\documentclass[iop,numberedappendix]{emulateapj}
\usepackage{graphicx}

\usepackage[english]{babel}

\usepackage{import}

\usepackage{afterpage}

\usepackage[plain]{fancyref}

\usepackage{gensymb}
\usepackage{amssymb}
\usepackage{amsmath}
\usepackage{bm}
\usepackage{mathptmx}
\usepackage{mathtools}
\DeclareMathAlphabet{\mathpzc}{OT1}{pzc}{m}{it}

\usepackage[usenames, dvipsnames]{color}
\usepackage{xcolor}

\usepackage{multirow}
\usepackage{booktabs}
\usepackage[flushleft]{threeparttable}
\usepackage{adjustbox}

\newcommand{\subsc}[1]{
_{\mbox{\tiny #1}}}

\newcommand{\supsc}[1]{
^{\mbox{\tiny \ #1}}}

\newcommand{\vect}[1]{\bm{#1}}
\newcommand{\matr}[1]{\bm{#1}}

\newcommand{\DCoffset}{\bar{T}\subsc{SZ}}

\usepackage[breaklinks,
            colorlinks,
            backref,
            citecolor=blue,
            plainpages=false]{hyperref}

\usepackage{natbib}
\begin{document}

\submitted{Submitted to the Astrophysical Journal: December 19, 2016}

\title{Constraints on the Mass, Concentration, and Nonthermal Pressure Support of Six CLASH Clusters from a Joint Analysis of X-ray, SZ, and Lensing Data}

\author{Seth~R.~Siegel\footnotemark[1,2,3]}
\footnotetext[1]{Division of Physics, Math, and Astronomy, California Institute of Technology, Pasadena, CA 91125}
\footnotetext[2]{Department of Physics, McGill University, 3600 Rue University, Montr\'eal, QC H3A 2T8, Canada}
\footnotetext[3]{\email{seth.siegel@mcgill.ca}}

\author{Jack~Sayers\footnotemark[1]}

\author{Andisheh~Mahdavi\footnotemark[4]}
\footnotetext[4]{Department of Physics and Astronomy, San Francisco State University, San Francisco, CA 94131}

\author{Megan~Donahue\footnotemark[5]}
\footnotetext[5]{Physics and Astronomy Department, Michigan State University, East Lansing, MI, 48824}

\author{Julian~Merten\footnotemark[6]}
\footnotetext[6]{Oxford University, Keble Road, Oxford OX1 3RH, United Kingdom}

\author{Adi~Zitrin\footnotemark[1,7]}
\footnotetext[7]{Hubble Postdoctoral Fellow}

\author{Massimo~Meneghetti\footnotemark[8]}
\footnotetext[8]{Jet Propulsion Laboratory, Pasadena, CA 91109}

\author{Keiichi~Umetsu\footnotemark[9]}
\footnotetext[9]{Institute of Astronomy and Astrophysics, Academia Sinica, P.O. Box 23-141, Taipei 10617, Taiwan}

\author{Nicole~G.~Czakon\footnotemark[9]}

\author{Sunil~R.~Golwala\footnotemark[1]}

\author{Marc~Postman\footnotemark[10]}
\footnotetext[10]{Space Telescope Science Institute, 3700 San Martin Drive, Baltimore, MD 21208}

\author{Patrick~M.~Koch\footnotemark[9]}

\author{Anton~M.~Koekemoer\footnotemark[10]}

\author{Kai-Yang~Lin\footnotemark[9]}

\author{Peter~Melchior\footnotemark[11,12]}
\footnotetext[11]{Center for Cosmology and Astro-Particle Physics, The Ohio State University, Columbus, OH 43210}
\footnotetext[12]{Department of Physics, The Ohio State University, Columbus, OH 43210}

\author{Sandor~M.~Molnar\footnotemark[9]}

\author{Leonidas~Moustakas\footnotemark[8]}

\author{Tony~K.~Mroczkowski\footnotemark[13]}
\footnotetext[13]{ESO - European Organization for Astronomical Research in the Southern hemisphere, Karl-Schwarzschild-Str\@. 2, D-85748 Garching b\@. M{\" u}nchen, Germany}

\author{Elena~Pierpaoli\footnotemark[14]}
\footnotetext[15]{University of Southern California, Los Angeles, CA 90089}

\author{Jennifer~Shitanishi\footnotemark[14]}

\begin{abstract}
We present a joint analysis of \emph{Chandra} X-ray observations, Bolocam thermal Sunyaev-Zel’dovich (SZ) effect observations, Hubble Space Telescope (HST) strong lensing data, and HST and Subaru Suprime-Cam weak lensing data.  The multiwavelength dataset is used to constrain parametric models for the distribution of dark and baryonic matter in a sample of six massive galaxy clusters selected from the Cluster Lensing And Supernova survey with Hubble (CLASH).  For five of the six clusters, the multiwavelength dataset is well described by a relatively simple model that assumes spherical symmetry, hydrostatic equilibrium, and entirely thermal pressure support.  The joint analysis yields considerably better constraints on the total mass and concentration of the cluster compared to analysis of any one dataset individually.  The subsample of five galaxy clusters is used to place an upper limit on the fraction of pressure support in the intracluster medium (ICM) due to nonthermal processes, such as turbulence and bulk flow of the gas.  We constrain the nonthermal pressure fraction at $r_{500c}$ to be $< 0.11$ at $95\%$ confidence.  This is in tension with state-of-the-art hydrodynamical simulations, which predict a nonthermal pressure fraction of $\approx 0.25$ at $r_{500c}$ for clusters of similar mass and redshift.  This tension may be explained by the sample selection and/or our assumption of spherical symmetry.
\end{abstract}



\section{Introduction}

Galaxy clusters play a unique role in the standard theory of structure formation as the largest objects to have undergone gravitational collapse.  This makes them a powerful tool for understanding the hierarchical process of structure formation and the cosmological backdrop in which it occurs.  Galaxy clusters have a wealth of observable properties through which they can be detected and studied.  They are populated with luminous galaxies that emit light in the optical and infrared regions of the spectrum.  They gravitationally lens the light emitted from background galaxies -- a process that is sensitive to the total cluster mass, the majority of which is attributed to dark matter.  Finally, they are pervaded by a diffuse, hot and ionized gas known as the intracluster medium (ICM) that accounts for the majority {\small ($\sim 90\%$)} of the baryonic mass.  The ICM emits X-rays through thermal bremsstrahlung radiation \citep{sarazin:1988} and inverse-Compton scatters Cosmic Microwave Background (CMB) photons through the thermal Sunyaev-Zel'dovich (SZ) effect \citep{sunyaev:1972}.

The hydrodynamical state of the ICM can be understood from analytical considerations, and numerical simulations can be used to make detailed predictions (e.g., \cite{shaw:2010,battaglia:2012,nelson:2014}).  However, it is not yet known how well these simulations account for many of the complicated but relevant baryonic processes that take place during cluster formation.  These processes include star formation, energy loss via radiative cooling, energy injection and metal enrichment via active galactic nuclei and supernovae winds, turbulence, and, at the cluster outskirts, incomplete virialization and bulk flow.  Our lack of knowledge is especially evident in the cluster outskirts, where there is sparse observational data to ground the predictions made by hydrodynamical simulations.

In our current understanding of cluster formation, an initial fast collapse is followed by a series of major mergers and the slow growth of the cluster outskirts through accretion of the surrounding intergalactic medium (IGM).  The cold IGM infalls at supersonic speeds and is shock heated near the virial radius.  The accretion shocks thermalize the majority of the kinetic energy acquired by the gas during infall.  Recent work suggests, however, that this mechanism does not result in complete virialization, and that some fraction of the kinetic energy remains in bulk and turbulent flow of the ICM \citep{cavaliere:2011a}.  These bulk and turbulent flows contribute to the support of the ICM against gravity, in the same was as thermal pressure, and thus such contributions are termed ``nonthermal pressure''.  Recent numerical simulations predict that the nonthermal component contributes $\sim40\%$ of the total pressure at the virial radius, with unrelaxed clusters showing systematically higher fractions compared to relaxed systems \citep{lau:2009,battaglia:2012,nelson:2014}.

Quantifying the level of nonthermal pressure support present in clusters can improve the constraining power of a number of cosmological probes.  As one example, the amplitude of the thermal SZ power spectrum is sensitive to several parameters in the standard $\Lambda \mbox{CDM}$ cosmology, primarily the amplitude of the initial density perturbations $\sigma_{8}$ \citep{komatsu:2002}.   However, in order to predict how much power one should expect to see in the thermal SZ signal for a given cosmology, one must know the level of nonthermal support in clusters over a wide range of masses and redshifts, and at large radii.  Current simulations differ in their treatment of the cluster thermal state and thus vary by as much as $60\%$ in their predictions of the thermal SZ amplitude \citep{shaw:2010,battaglia:2010,trac:2011,mccarthy:2014}.  As a result, theoretical modeling uncertainties limit the constraint on $\sigma_{8}$ that can be derived from such measurements \citep{reichardt:2012}, offering only marginal improvements over the existing constraint from measurements of the CMB, $\mbox{H}_{0}$, and BAO.

Constraints on the level of nonthermal pressure support present in individual clusters can be obtained by combining multiple observables in a joint analysis.  The magnitude of the SZ temperature decrement measured at a particular frequency is proportional to the thermal electron pressure $P_{e} = k_{B} n_{e} T_{e}$ integrated along the line of sight.  X-ray surface brightness is proportional to the integral of a slightly different combination of the number density and temperature --- specifically $n_{e}^{2} T_{e}^{1/2}$ --- and thus provides a slightly different probe of the ICM thermal state.  The gravitational lensing of background galaxies is sensitive to the total mass density projected along the line-of-sight, and is independent of the thermal state of the ICM.  The difference between the thermal pressure inferred from X-ray/SZ observations and the total pressure gradient necessary to balance the gravitational force inferred from lensing observations provides a measurement of the nonthermal pressure support.  

Joint analysis of X-ray, SZ, and lensing data has been published for Abell~1835 by \citet{morandi:2012} and for Abell~1689 by \citet{sereno:2013} and \citet{umetsu:2015a}.  Both Abell~1835 and Abell~1689 are high-mass clusters ($M\subsc{200c} > 1 \times 10^{15} \ \mbox{M}_{\odot}$) at relatively low redshift ($z < 0.3$).  While the three analyses differ in implementation, they share a set of assumptions that underly the modeling of the multiwavelength dataset.  All assume parametric functions in a triaxial coordinate system for the dark matter and ICM profiles, the latter given enough freedom to constrain the full thermodynamics and potential nonthermal motion of the ICM.  In the case of Abell~1835, the nonthermal pressure fraction increases with radius to a value of $0.18 \pm 0.07$ at the outer edge of the cluster.  In the case of Abell~1689, both \citet{sereno:2013} and \citet{umetsu:2015a} find nonthermal pressure fractions that are relatively constant with radius at a value of $0.4 \pm 0.1$.  Numerical simulations suggest that galaxy clusters have little nonthermal pressure at small radii outside of the central core, but that the nonthermal pressure fraction rises monotonically to $\simeq 0.4$ at the outskirts.  Consequently, Abell~1835 shows a lower than expected nonthermal pressure fraction, whereas Abell~1689 shows a larger than expected nonthermal pressure at small and intermediate radii.  In order to provide comprehensive observational constraints, the techniques established in these works must be applied to a large sample of galaxy clusters with well-defined selection criteria.

\setlength{\textfloatsep}{2mm}

\begin{deluxetable*}{ l c c c c c c c c c }
\tablecolumns{9}
\tablecaption{Characteristics of the multiwavelength observations of the six galaxy clusters in our sample. \label{tab:spherical_sample}}
\tablehead{Name & 
           \colhead{$z$} & 
           \colhead{RA} 		& 
           \colhead{DEC} 		& 
           \colhead{SZ S/N}     & 
           \colhead{\emph{Chandra}}     & 
           \colhead{$N\subsc{sys}$\,\tablenotemark{a}} & 
           \colhead{HST $\rho\subsc{gal}$\,\tablenotemark{b}} & 
           \colhead{Subaru $\rho\subsc{gal}$\,\tablenotemark{c}} \\
           \colhead{}            &
           \colhead{} 	         &	   
           \colhead{(J2000)} 	 & 
           \colhead{(J2000)}	 &		 
           \colhead{}            & 
           \colhead{Time (ksec)} &                
           \colhead{}            &  
           \colhead{($\mbox{arcmin}^{-2}$)} & 
           \colhead{($\mbox{arcmin}^{-2}$)}}
\startdata
Abell 383			& 0.187 & 02:48:03.40 &	 -03:31:44.9 &	9.6  & 38.8  & 9 & 50.7 & 9.0 \\
Abell 611			& 0.288 & 08:00:56.82 &	 +36:03:23.6 &	10.8 & 36.1  & 4 & 42.3 & 8.8 \\
MACS J0429.6-0253	& 0.399 & 04:29:36.05 &	 -02:53:06.1 &	8.9  & 23.2  & 3 & 42.4 & 12.0 \\ 
MACS J1311.0-0310	& 0.494 & 13:11:01.80 &	 -03:10:39.8 &	9.6  & 63.2  & 2 & 33.7 & 20.2 \\
MACS J1423.8+2404	& 0.545 & 14:23:47.88 &	 +24:04:42.5 &	9.4  & 115.6 & 5 & 75.3 & 9.8 \\
MACS J1532.8+3021	& 0.363 & 15:32:53.78 &	 +30:20:59.4 &	8.0  & 89.0  & 0 & 35.9 & 16.6
\enddata
\tablenotetext{a}{The number of multiple-image systems used in the strong lensing analysis of \citet{merten:2015}.}
\tablenotetext{b}{The surface-number density of background selected galaxies in the HST field used for the weak lensing analysis of \citet{merten:2015}.}
\tablenotetext{c}{The surface-number density of background selected galaxies in the Subaru field used for the weak lensing analysis of \citet{merten:2015} and derived from the work of \citet{umetsu:2014}.}
\end{deluxetable*}

In this paper, we fit parametric models to the combined X-ray, SZ, and lensing data available for a subset of 6 clusters selected from the Cluster Lensing and Supernova survey with Hubble (CLASH) sample \citep{postman:2012}.  By doing so, we are able to achieve significant improvements in the constraints on the distribution of dark and baryonic matter compared to single-probe analyses.  We are also able to place an upper limit on the level of nonthermal pressure support.

\Fref{sec:cluster_sample} introduces the sample of 6 clusters that are the topic of this paper and describes their selection criteria.  \Fref{sec:jaco_model} presents the theoretical model used to describe the multiwavelength dataset.  \Fref{sec:jaco_data_description} provides an overview of the data used in this analysis, including \emph{Chandra} observations of the X-ray emission, Bolocam observations of the SZ effect, \emph{Hubble Space Telescope} (HST) gravitational strong lensing measurements, and HST and Subaru weak lensing measurements.  \Fref{sec:jaco_method} describes the specifics of our analysis: the fitting method employed and the determination of the optimal model.  \Fref{sec:jaco_results} presents the results.  \Fref{sec:jaco_discussion} concludes with a discussion of the results.  Throughout this work we assume a $\Lambda \mbox{CDM}$ cosmology with $\Omega_{m} = 0.30$, $\Omega_{\Lambda} = 0.70$, and $h \equiv 0.70 \times h_{70} = 0.70$, where $H_{0} = h \times 100 \ \mbox{km} \ \mbox{s}^{-1} \ \mbox{Mpc}^{-1}$.

\section{Cluster Sample}
\label{sec:cluster_sample}

The CLASH sample consists of 25 galaxy clusters that cover a significant portion of cluster formation history $(0.2 < z < 0.9)$ and span almost an order of magnitude in mass $(5$--$20 \times 10^{14} \ \mbox{M}_{\odot})$ \citep{postman:2012}.  The multiwavelength dataset available for the CLASH sample is unparalleled in terms of its breadth and quality.  All 25 clusters have been observed with the \emph{Chandra X-ray Observatory} and 15 of the clusters also have \emph{XMM-Newton} data available.  The thermal SZ effect has been measured at 140 GHz for all 25 of the clusters with Bolocam, a millimeter-wave imaging camera at the Caltech Submillimeter Observatory (CSO).  HST provides 16-band, high-precision strong lensing data in the cluster core and weak lensing data at intermediate radii, while the multi-band Suprime-Cam on the Subaru Telescope provides wide-field weak lensing data of the outskirts, thereby characterizing the total matter distribution over a wide range of scales.

The CLASH sample was selected based on one of two selection criteria.  Twenty were selected from X-ray based compilations of massive, dynamically relaxed galaxy clusters with the primary criteria being a highly regular X-ray morphology.  More specifically, these clusters have \emph{Chandra} X-ray surface brightness images that consist of a single, well-defined peak and round, concentric isophotes.  The other five clusters were selected because they have large Einstein radii and thus are exceptionally strong gravitational lenses \citep{postman:2012}.

This paper is intended to act as a proof-of-principle that the rich multiwavelength dataset that now exists for each CLASH cluster can be understood in the context of a relatively simple parametric model, and explore how the different data come together to constrain this model.  For this initial demonstration we assume a spherically symmetric model and focus on a subset of CLASH clusters that have round and regular morphologies in both X-ray and SZ maps.  We emphasize that a round and regular morphology is a necessary, but not sufficient, condition for our assumed spherical model to provide an accurate description of the cluster. For example, objects that appear round in the plane of the sky are often elongated along the line of sight, due to the fact that massive clusters tend to have a prolate geometry \citep{meneghetti:2010, rasia:2012, meneghetti:2014}. As detailed in Section~\ref{sec:jaco_results}, such an elongation could potentially bias some of the constraints we derive using a spherical model. However, for all but one cluster in our study, the spherical model provides an adequate fit to the data, implying that any elongation bias is subdominant to the statistical uncertainties.

The cluster subset for this analysis is chosen in the following way.  We start by restricting our attention to the 20 CLASH clusters that were chosen based on X-ray morphology.  Simulations suggest that these 20 clusters are predominately relaxed ($\sim 70\%$) and largely free of orientation bias \citep{meneghetti:2014}.  A cluster must satisfy two additional requirements in order to be placed in our sample.  First, the SZ morphology must be circular.  This requirement is implemented by fitting the SZ image alone using circular and elliptical versions of the generalized-NFW model (gNFW) for the thermal pressure \citep{nagai:2007a,arnaud:2010}, and examining whether the elliptical model is preferred by performing a statistical $F$-test.  \citet{czakon:2015} outlines this procedure and presents the results for all CLASH clusters.   Second, we require that the X-ray centroid shift parameter, $w_{500c}$, is less than 0.006.  The centroid shift parameter is the standard deviation in units of $r_{500c}$ of the separation between the peak and centroid of the X-ray emission calculated in increasing aperture sizes up to $r_{500c}$.  The $w_{500c}$ values for all CLASH clusters were calculated using \emph{Chandra} data according to the procedure described in \citet{maughan:2008,maughan:2012} and are presented in \citet{sayers:2013}.

Of the 20 X-ray selected CLASH clusters, 8 satisfy both requirements.  However, a qualitative comparison of the mass profiles obtained from independent analyses of the gravitational lensing data by \citet{merten:2015} and \citet{umetsu:2015} suggested possible discrepancies for 2 of the 8 clusters: MACS~J1931.8-2634 and MS~2137.3-2353.  Since we were not confident in the lensing constraints for these two clusters at the time of the analysis, we removed them from our sample.  Note that \citet{merten:2015} performed a joint analysis of HST strong lensing and HST/Subaru weak lensing shear data, whereas \citet{umetsu:2015} also included HST/Subaru weak lensing magnification data.  In the case of MACS~J1931.8-2634, the discrepancy is likely due to unaccounted systematic uncertainties in the calibration of the magnification data for clusters at low galactic latitude.  In the case of MS~2137.3-2353, a quantitative comparison has since demonstrated that the two analyses are indeed consistent within their respective uncertainties \citep{umetsu:2015}.

\Fref{tab:spherical_sample} lists the 6 CLASH clusters that make up our sample, presents their basic properties, and provides metrics for the quality of their observations.

\section{Cluster Model}
\label{sec:jaco_model}

We assume that the galaxy cluster is spherically symmetric and use parametric functions to describe the radial dependence of the total matter density, gas density, metallicity, and fraction of the total pressure support sourced by nonthermal processes.  By further assuming that the cluster is in a state of hydrostatic equilibrium, we can predict all observable quantities of interest.

\subsection{Total Matter Density}

We model the total matter density with the Navarro-Frenk-White profile (NFW hereafter) \citep{navarro:1995,navarro:1996}
\begin{align}
\rho\subsc{tot}(r) & = \rho\subsc{tot,0}\left(\frac{r}{r_{s}}\right)^{-1} \left(1 + \frac{r}{r_{s}}\right)^{-2} \ ,
\end{align}
which is defined by two parameters: a normalization $\rho\subsc{tot,0}$ and scale radius $r_{s}$.  It is standard to reparameterize in terms of the total mass and concentration at a particular overdensity radius
\begin{align}
M_{\mbox{\tiny tot, } \Delta \mbox{\tiny ref}} & \equiv 4 \pi r_{s}^3 \rho\subsc{tot,0} \left[\ln{\left(\frac{r_{s} + r_{\Delta \mbox{\tiny ref}}}{r_s}\right)} - \frac{r_{\Delta \mbox{\tiny ref}}}{r_{s} + r_{\Delta \mbox{\tiny ref}}}\right] \\
c_{\Delta \mbox{\tiny ref}} & \equiv \frac{r_{\Delta \mbox{\tiny ref}}}{r_{s}} \ ,
\end{align}
where $r_{\Delta \mbox{\tiny ref}}$ denotes the radius at which the average enclosed density is $\Delta$ times some reference density.  Two common reference densities that we will employ in this work are the critical density of the universe and the mean matter density of the universe
\begin{align}
\rho_{c}(z) & = \frac{3 H_{0}^2}{8 \pi G} \left[ \Omega_{m} (1 + z)^3 + \Omega_{\Lambda} \right] , \\[+2mm]
\rho_{m}(z) & = \frac{3 H_{0}^2}{8 \pi G} \Omega_{m} (1 + z)^3 \ .
\end{align}
The overdensity radius $r_{\Delta \mbox{\tiny ref}}$ is determined by solving the implicit equation
\begin{align}
M_{\mbox{\tiny tot, }\Delta \mbox{\tiny ref}} & = \frac{4}{3} \pi r_{\Delta \mbox{\tiny ref}}^3 \Delta \rho\subsc{ref} \ .
\end{align}
Common overdensity radii that are used throughout the literature and will be referenced in this paper are $r_{2500c} < r_{500c} < r_{200c}   < r_{200m}$.

\subsection{Gas Density}
We model the gas density as
\begin{align}
\label{eq:gas_density}
\rho\subsc{gas}(r)  & =  \rho\subsc{gas,0} \left(\frac{r}{r\subsc{gas}}\right)^{-\alpha} \left(1 + \left(\frac{r}{r\subsc{gas}}\right)^{2}\right)^{\left(\alpha - 3 \beta\right) / 2} \nonumber \\
& \phantom{= \rho\subsc{gas,0} \left(\frac{r}{r\subsc{gas}}\right)} \times \left(1 + \left(\frac{r}{r\subsc{gas,\ outer}}\right)^{\delta}\right)^{-\epsilon / \delta} \nonumber \\
& \phantom{ = } + \rho\subsc{gas,\ core} \left(1 + \left(\frac{r}{r\subsc{gas,\ core}}\right)^{2}\right)^{-3 \beta\subsc{core} / 2} \ ,
\end{align}
which is inspired by the expression used in \citet{vikhlinin:2006a} to describe the X-ray surface brightness of nearby relaxed galaxy clusters.  \Fref{eq:gas_density} is the sum of two $\beta$-models \citep{cavaliere:1978}, with the first $\beta$-model modified by two additional factors.  The $r^{-\alpha}$ power-law factor allows for a central cusp instead of the flat core inherent to the $\beta$-model \citep{pratt:2002}.  This is necessary to describe cool-core clusters, which tend to exhibit a nonzero logarithmic slope $\alpha \approx 0.5$ in the cluster core \citep{sanderson:2010}.  The $r^{-\epsilon}$ factor allows for the logarithmic slope of the gas density to steepen by some amount $\epsilon$ at radius $r\subsc{gas,outer}$ (with $r\subsc{gas,outer} > r\subsc{gas}$).  The parameter $\delta$ controls how quickly the gas density transitions from the $r^{-3 \beta}$ power-law to the $r^{-3 \beta - \epsilon}$ power-law; we fix $\delta = 4$ for this analysis.  Steepening of the gas density profile in the cluster outskirts is observed in hydrodynamical simulations \citep{roncarelli:2006}, X-ray observations of individual clusters \citep{vikhlinin:1999,neumann:2005,vikhlinin:2006a,croston:2008,sanderson:2010}, and the stacked analysis of X-ray data from many clusters \citep{morandi:2015}.  The second $\beta$-model aids in the description of the core region of the cluster.  To ensure this role, we force $r\subsc{gas,core} < 50 \ \mbox{kpc}$ and fix $\beta\subsc{core} = 1$.  We note that our model differs from that presented in \citet{vikhlinin:2006a} in two regards.  First, we assume a value $\delta = 4$ resulting in a slightly more rapid transition than the \citet{vikhlinin:2006a} model, which assumes a value $\delta = 3$.  This choice was motivated by a similar multiwavelength analysis performed by \citet{morandi:2012}.  Second, we model the gas density $\rho\subsc{gas}$ whereas they model the X-ray surface brightness, which is proportional to $\rho\subsc{gas}^2$.  Therefore, our prediction for the X-ray surface brightness will have a cross-term between the first and second $\beta$-model that is not present in their model.  This will result in slightly different gas density profiles for the same set of parameter values in the region where the core $\beta$-model transitions to the primary $\beta$-model.

\subsection{Nonthermal Pressure Support}

We assume that the total pressure is the sum of the thermal pressure and the nonthermal pressure
\begin{align}
P\subsc{tot} & = P\subsc{th} + P\subsc{nth} \\[+2mm]
& = \frac{k_{B} T \rho\subsc{gas}}{\mu m_{p}} + P\subsc{nth} \ ,
\end{align}
where $m_{p}$ is the proton mass, $\mu$ is the mean molecular weight of the ICM, and $T$ is the temperature of the ICM.  We model the nonthermal pressure fraction as
\begin{align}
\frac{P\subsc{nth}}{P\subsc{tot}}\left(r\right) \equiv \mathpzc{F}(r) & = \mathpzc{F}\subsc{outer}(r) + \mathpzc{F}\subsc{inner}(r)
\end{align}
with
\begin{align}
\mathpzc{F}\subsc{outer}(r) & = C \left\{1 - A \left(1 + \exp{\left[\left(\frac{r / r_{200m}}{B}\right)^{\gamma}\right]}\right) \right\}
\end{align}
and
\begin{align}
\mathpzc{F}\subsc{inner}(r) & = D \left(1 + \left(\frac{r / r_{200m}}{E}\right)^{4} \right)^{-\zeta/4} \ .
\end{align}
The $\mathpzc{F}\subsc{outer}$ term is a scaled version of the \citet{nelson:2014} empirical fitting formula used to describe the mean nonthermal pressure fraction observed in the region $r \gtrsim 0.1 \times r_{200,m}$ in a mass-limited sample of clusters from a high-resolution hydrodynamical simulation.  We fix the radial dependence to that observed in the simulation by fixing the parameters $\left[A, \ B, \ \gamma \right]$ at the \citet{nelson:2014} best-fit values $\left[0.452, \ 0.841, \ 1.628\right]$, and allow only the normalization $C$ to float.  The $\mathpzc{F}\subsc{inner}$ term allows the nonthermal pressure fraction to increase by some amount $D$ in the cluster core.  We require that $E < 0.1$, which ensures that this inner term only describes regions interior to those examined in the simulations, which are well described by $\mathpzc{F}\subsc{outer}$.  There are a number of physical processes that can strongly influence the thermodynamic state of the ICM in the cluster core.  Our goal in introducing the second term is to decouple the nonthermal pressure in the outer regions of the cluster, which is the quantity we would like to constrain, from that in the core.

We assume that the ICM is in a state of equilibrium where the inward gravitational pull is balanced by a pressure gradient.  This assumption of hydrostatic equilibrium is expressed as the following differential equation:
\begin{align}
\label{eq:hydrostatic_equilibrium}
\nabla P\subsc{tot} & = -\rho\subsc{gas} \nabla \Phi \ ,
\end{align}
where $\Phi$ is the gravitational potential.  We note that \Fref{eq:hydrostatic_equilibrium} contains nonthermal pressure support as part of $P\subsc{tot}$, and it therefore differs from the standard definition of hydrostatic equilibrium that is commonly used in the literature and implies entirely thermal pressure support.  We are allowing a nonthermal pressure component sourced by bulk and turbulent motions of the gas to provide some fraction of the support necessary to prevent gravitational collapse.  For our model, \Fref{eq:hydrostatic_equilibrium} is written as
\begin{align}
& \frac{d}{dr}\left[\frac{1}{1 - \mathpzc{F}(r)}\frac{\rho\subsc{gas}(r)k_{B}T(r)}{\mu m_{p}}\right] = - \frac{G M\subsc{tot}(r) \rho\subsc{gas}(r)}{r^2} \ ,
\end{align}
where $G$ is the gravitational constant and $k_{B}$ is the Boltzmann constant.  Integration yields
\begin{align}
& k_{B} T(r) = k_{B} T\subsc{trunc} + \nonumber \\
& \phantom{k_{B} T(r) = } \left(1 - \mathpzc{F}(r)\right) \frac{\mu m_{p}}{\rho\subsc{gas}(r)} \int_{r}^{r\subsc{trunc}} \frac{G M\subsc{tot}(x)\rho\subsc{gas}(x)}{x^{2}} dx \ ,
\end{align}
where $T\subsc{trunc}$ is the temperature at some radius $r\subsc{trunc}$ that designates the outer boundary of the ICM.  Our model does not assume an explicit parameterization for the temperature, rather it is an internal variable that is derived from the total density, gas density, and nonthermal pressure fraction assuming hydrostatic equilibrium.

We must the model the metallicity of the ICM because it influences the X-ray cooling function and thus the X-ray emission.  We describe the metallicity with the function
\begin{align}
\label{eq:metallicity}
Z(r) = Z_{0} \left(1 + \left(\frac{r}{r_{Z}}\right)^2\right)^{-3 \beta_{Z}/2} \ ,
\end{align}
which allows for a central metallicity $Z_{0}$ that transitions to a power-law $r^{-3 \beta_{Z}}$ at radius $r_{Z}$ \citep{pizzolato:2003}.  The electron and hydrogen number density are given by
\begin{align}
n_{H}(r) & = \frac{X}{m_{p}} \rho\subsc{gas}(r) \ , & 
n_{e}(r) & = \left<\frac{n_{e}}{n_{H}}\right> n_{H}(r)  \ ,
\end{align}
where $X$ denotes the hydrogen mass fraction and $<n_{e} / n_{H}>$ the ion to hydrogen ratio.  The mean molecular weight $\mu$, which appears in several equations above, along with $X$ and $<n_{e} / n_{H}>$, are mild functions of the metallicity, and are calculated using an absolute metallicity given by \Fref{eq:metallicity} with the relative abundances fixed on the photospheric values given by \citet{grevesse:1998}.

\subsection{Observables}

All observable quantities of interest can be predicted from the above model.  Let $D_{A}(z)$ denote the angular diameter distance, $\theta$ the angular separation from the cluster center,  and $R = D_{A} \theta$ the radius from the cluster center projected on the plane of the sky.  

\subsubsection{X-ray}

The X-ray flux from the cluster measured at an energy $h\nu$ within an annulus of inner radius $R_{1}$ and outer radius $R_{2}$ is given by
\begin{align}
\label{eq:xray_proj}
S & = \frac{1}{4 \pi D_{L}^2}\int_{R_{1}}^{R_{2}} 2 \pi R dR \nonumber \\
& \phantom{ = \frac{1}{4 \pi D_{L}^2}} \int_{R}^{r\subsc{trunc}} n_{e}(r) n_{H}(r) \Lambda\left[h\nu', T(r), Z(r)\right] \frac{2 r dr}{\sqrt{r^2 - R^2}} \ ,
\end{align}
where $D_{L}(z)$ is the luminosity distance, $h\nu' = h\nu / (1+ z)$ is the energy in the cluster rest frame, and $\Lambda\left[h\nu', T(r), Z(r)\right]$ is the X-ray cooling function.  In addition to the X-ray flux from the cluster our model includes X-ray flux from a uniform thermal background:
\begin{align}
S\subsc{sbkg} = A\subsc{sbkg} \ \Lambda\left[\nu, T\subsc{sbkg}, Z_{\odot} \right] \ .
\end{align}
This accounts for galactic soft X-ray emission which varies across the sky and therefore is not adequately subtracted using a background observation (see \citealt{mahdavi:2007} for more details).  Here $A\subsc{sbkg}$ acts as an overall normalization and $T\subsc{sbkg} \sim 0.5 \ \mbox{keV}$ is the temperature of the galactic, X-ray emitting gas.

\subsubsection{Thermal SZ Effect}

The thermal SZ effect results in a distortion of the CMB blackbody spectrum.  The change in the temperature of the CMB measured at a frequency $\nu$ and projected radius $R$ is given by
\begin{align}
\label{eq:sz_proj1}
T\subsc{SZ} & = T\subsc{CMB} f(x) y \ .
\end{align}
The function $f(x)$ encodes the frequency dependence of the classical distortion
\begin{align}
\label{eq:sz_proj2}
f(x) = x \frac{e^{x} + 1}{e^{x} - 1} - 4 \ ,
\end{align}
where $x \equiv h \nu / k_{B} T\subsc{CMB}$.  The Compton $y$ parameter sets the magnitude of the distortion and is proportional to the integral of the thermal electron pressure along the line of sight
\begin{align}
\label{eq:sz_proj3}
y = \frac{\sigma_{T}}{m_{e} c^2}\int_{R}^{r\subsc{trunc}} n_{e}(r) k_{B} T(r)  \left[1 + \delta_{R}(x, T(r)) \right] \frac{2 r dr}{\sqrt{r^2 - R^2}} \ ,
\end{align}
where $\sigma_{T}$ is the Thomson cross section, $c$ is the speed of light, and $m_{e}$ is the mass of the electron.  The quantity $\delta_{R}(x, T(r))$ is a correction for the relativistic motion of the electrons, which we approximate using the expansion given in \citet{itoh:1998}. 

\subsubsection{Gravitational Lensing}

Based on the generally applicable assumption that the line of sight extent of the mass distribution is small compared to the distances between the observer, mass distribution, and background galaxies, gravitational lensing of the light from those galaxies is described by a lens equation $\vect{\beta} = \vect{\theta} - \vect{\alpha}(\vect{\theta})$ which maps the angular coordinates of the galaxy in the source plane $\vect{\beta} = [\beta_{1},\beta_{2}]$ to the coordinates in the lens plane $\vect{\theta} = [\theta_{1},\theta_{2}]$ through a deflection angle $\vect{\alpha} = [\alpha_{1},\alpha_{2}]$ (see, e.g., \citealt{bartelmann:2001, bartelmann:2010}).  We can define a lensing potential
\begin{align}
\Psi(\vect{\theta}) = \frac{D_{ls}}{D_{l}D_{s}} \frac{2}{c^2} \int_{-\infty}^{\infty} \Phi(R,\ell) d\ell \ ,
\end{align}
which is just the three-dimensional gravitational potential projected along the line of sight and rescaled.  In the above equation $D_{s}$, $D_{l}$, and $D_{ls}$ denote the observer-source, observer-lens, and lens-source angular diameter distances, respectively.  The deflection angle is then equal to the gradient of the lensing potential
\begin{align}
\vect{\alpha}(\vect{\theta}) & = \nabla \Psi(\vect{\theta}) \ .
\end{align}
The convergence $\kappa$ and complex shear $\vect{\gamma} = [\gamma_{1}, \gamma_{2}]$ of the lens are also related to the lensing potential through the equations
\begin{align}
\kappa(\vect{\theta}) & = \frac{1}{2} \left(\frac{\partial^2}{\partial \theta_{1}^2} + \frac{\partial^2}{\partial \theta_{2}^2} \right) \Psi(\vect{\theta}) = \frac{\Sigma(\vect{\theta})}{\Sigma\subsc{crit}} \label{eq:lensing_convergence} \\[+2mm]
\gamma_{1}(\vect{\theta}) & = \frac{1}{2} \left(\frac{\partial^2}{\partial \theta_{1}^2} - \frac{\partial^2}{\partial \theta_{2}^2} \right) \Psi(\vect{\theta}) \\[+2mm]
\gamma_{2}(\vect{\theta}) & = \frac{\partial}{\partial\theta_{1}}\frac{\partial}{\partial\theta_{2}} \Psi(\vect{\theta}) \ .
\end{align}
Here $\Sigma(\vect{\theta})$ is the surface mass density and $\Sigma\subsc{crit}$ is the critical surface mass density for lensing, given by
\begin{align}
\Sigma\subsc{crit} & = \frac{c^2}{4 \pi G}\frac{D_{s}}{D_{ls}D_{l}} \ ,
\end{align}
where $G$ is the gravitational constant.

In the weak lensing regime the gravitational shear introduces a complex ellipticity $\vect{e}$ to the images of background galaxies which is approximately equal to $\vect{\gamma}$ and is described by the reduced shear
\begin{align}
\langle \vect{e} \rangle = \frac{\vect{\gamma}}{1 - \kappa} \ ,
\end{align}
where $\langle \vect{e} \rangle$ denotes a local average necessary to mitigate the intrinsic ellipticity of the galaxies.  In the strong lensing regime, where multiple solutions to the lens equation are possible, more than one image of a single source can be observed.  These multiple images straddle critical lines whose locations are set by the relation
\begin{align}
(1 - \kappa)^{2} - \gamma^{2} = 0 \ .
\end{align}
The combined strong and weak lensing analysis outlined in the following section employs the location of the critical lines and the ellipticity of background galaxies to measure the convergence of the galaxy cluster.  According to our model the convergence measured at a projected radius $R$ is given by
\begin{align}
\label{eq:lensing_proj}
\kappa & = \frac{1}{\Sigma\subsc{crit}}\int_{R}^{\infty} \rho\subsc{tot}(r) \frac{2 r dr}{\sqrt{r^2 - R^2}} \ .
\end{align}

\section{Description of the Multiwavelength Dataset}
\label{sec:jaco_data_description}

\subsection{Chandra X-ray}

The reduction of the CLASH X-ray data is described in detail in \citet{donahue:2014} and we briefly summarize the procedure below.  The data is processed using CIAO 4.6.1 (released February 2014) and CALDB 4.5.9 (released November 2013).  Flares are identified as time intervals with outlier event rates in $0.5$--$7.0 \ \mbox{keV}$ light curves extracted from source-free areas of the detector.  Events coincident with a flare are removed from the event lists.  Bright point sources are identified using the CIAO \emph{wavdetect} algorithm and a map of the PSF size as a function of location on the detector.  Regions near the bright point sources are filtered from the event lists.  Each dataset is matched to a deep background file from a similar observation epoch, which is used to subtract contamination from faint point sources, galactic soft X-ray emission, and non-flaring particle events \citep{hickox:2007,markevitch:2003}.  The background files are filtered, reprojected, and rescaled to match the target observation.  The rescaling is done by adjusting the exposure time on the deep background file so that the event rate between $10-12 \ \mbox{keV}$ is equal to that in the cluster field.  This particular energy range is chosen because the effective area for X-ray photons is low and the event rate is dominated by high-energy particle events.

X-ray spectra are generated in concentric annular bins centered on the coordinates given in \Fref{tab:spherical_sample}.  The boundaries of the bins are selected so that at least 1500 photon counts from the cluster are contained in each annulus and the width of each annulus is at least a few times the width of the PSF.  Compared to the analysis of \citet{donahue:2014}, we have added one additional annulus to each cluster.  This annulus is located beyond the radius of the outermost annulus used in that work.  The spectra are binned in energy from $0.5$--$11.0 \ \mbox{keV}$ with a bin width of $38 \ \mbox{eV}$.  The same binning scheme is applied to both the observation file and the deep background file.  The individual weighted redistribution matrix file (RMFs) and ancillary response file (ARFs) are then computed.  The cluster field spectra $\vect{S\supsc{obs}}$, deep background spectra $\vect{S\supsc{bkg}}$, RMFs, and ARFs are all input to the multiwavelength analysis.

The spectra generated from the deep background file are eventually subtracted from the spectra generated from the target observation file.  Consider the energy bin $h\nu_{j}$ and the annulus with inner radius $R_{i}$ and outer radius $R_{i+1}$.  The resulting X-ray measurement is
\begin{align}
S_{ij} = S\supsc{obs}_{ij} - S\supsc{bkg}_{ij}
\end{align}
and the associated Poisson error is
\begin{align}
\sigma_{S_{ij}} = \sqrt{S\supsc{obs}_{ij} + S\supsc{bkg}_{ij}}
\end{align}
with units of $\mbox{counts} \ \mbox{sec}^{-1} \ \mbox{keV}^{-1}$.

\subsection{Bolocam Thermal SZ Effect}
\label{sec:bolocam_sz}

The thermal SZ effect has been measured at 140 GHz for the six clusters in our sample using Bolocam, a 144-element bolometric imaging camera at the Caltech Submillimeter Observatory \citep{glenn:1998,haig:2004}.  Bolocam has an $8 \ \mbox{arcmin}$ diameter circular field of view (FOV) and a $58 \ \mbox{arcsec}$ full width at half maximum point spread function (PSF).  The measurements were made over the course of 14 observing runs between 2006 and 2012 as part of a larger campaign that resulted in the creation of the Bolocam X-ray SZ (BOXSZ) sample of 45 galaxy clusters \citep{sayers:2013a,czakon:2015}.  We summarize the general properties of the SZ data products here, and direct the interested reader to \citet{sayers:2011} for a description of the data reduction, flux calibration, and noise estimation, and \citet{czakon:2015} for a description of the BOXSZ sample.  The SZ data products for all of the clusters in the BOXSZ sample are publicly available.~\footnote{\url{http://irsa.ipac.caltech.edu/data/Planck/release_2/ancillary-data/bolocam/}}

Noise sourced by fluctuations in atmospheric emission dominates the raw detector timestreams at long timescales.  The atmospheric noise is mitigated by subtracting the response-weighted mean detector signal and applying a $250 \ \mbox{mHz}$ high-pass filter \citep{sayers:2011}.  This data processing attenuates the cluster signal in a way that is mildly dependent on the cluster shape and also results in the loss of the image's mean signal.  To account for the attenuation of the cluster signal, a complex-valued two-dimensional map space Fourier transfer function is calibrated for each cluster.  The mean signal of the image is included as a free parameter $\DCoffset$ in our model fits.

Non-astronomical noise is estimated from 1000 jackknife realizations of the cluster image.  To account for astronomical noise sourced by CMB anisotropies and unresolved point sources, Gaussian random realizations of the $140 \ \mbox{GHz}$ sky are generated from SPT power spectrum measurements \citep{keisler:2011,reichardt:2012}, passed through the data processing pipeline, and added to each of the 1000 jackknife realizations.  Note that the SPT power spectrum measurements cover the full range of angular scales probed by the Bolocam images.  Known radio point sources have been subtracted from the Bolocam images, and random realizations of the estimated residual from the subtraction are injected into the each of the 1000 jackknife realizations as well.  It has been confirmed that the resulting 1000 noise realizations are statistically indistinguishable from observations of blank sky \citep{sayers:2011}.

The pixel-to-pixel covariance matrix of the SZ image is estimated as
\[ (\matr{C_{T\subsc{SZ}}})_{ij} = 
\begin{cases}
\dfrac{\left(\mbox{sensitivity}\right)^2}{t_{i}} &  i = j \\
\hfil 0 & i \neq j \ ,
\end{cases}
\]
where $t_{i}$ is the (known) integration time for pixel $i$.  The sensitivity is determined by fitting a Gaussian to a histogram of the product of the pixel value and the square root of the pixel integration time for all pixels in all 1000 noise realizations.  The assumption that the off-diagonal elements are zero is a good but not perfect description of the data.  The set of observations do not contain enough information to estimate the off-diagonal elements of the covariance matrix, and simplifying assumptions about the structure of the covariance matrix (e.g., that it is only a function of pixel separation) have proven false.  Instead, we carry out a test (described in \Fref{sec:sz_covariance}) to determine what effect the small inter-pixel correlations in the SZ image have on the resulting parameter constraints.  We find that the effect is not significant, and therefore ignore the off-diagonal noise terms throughout our analysis.  We also note that \citet{sayers:2011} demonstrates that the distribution of $\chi^2$ values obtained from fitting a model to the Bolocam SZ data accounting for inter-pixel correlations using the noise realizations is nearly identical to the theoretical $\chi^2$ distribution for the diagonal covariance matrix assumption.

The SZ images are $14 \ \mbox{arcmin} \times 14 \ \mbox{arcmin}$ with $20 \ \mbox{arcsec}$ square pixels.  For our analysis we only fit pixels with an angular separation $\theta \leq 6.33 \ \mbox{arcmin}$ from the center of the image.  This is the largest aperture wherein all pixels have an integration time $t > 0.25 \times t\subsc{max}$, where $t\subsc{max}$ is the maximum integration time achieved in the center of the image.  The input to the multiwavelength analysis is the image $\vect{T\subsc{SZ}}$ in units of $\mu K\subsc{CMB}$, the diagonal covariance matrix $\matr{C_{T\subsc{SZ}}}$, and the transfer function of the data processing pipeline.

\subsection{HST and Subaru Gravitational Lensing}  

The vast majority of the CLASH clusters have HST strong lensing, HST weak lensing, and Subaru Suprime-Cam weak lensing constraints.  \citet{merten:2015} outlines the procedure used to self-consistently combine these constraints into a nonparametric estimate of the lensing convergence profile.  We summarize the main steps of this procedure.

The strong lensing reduction begins by identifying multiple-image systems in the 16-band HST images using the method outlined in  \citet{zitrin:2009,zitrin:2015}.  The redshift associated to each multiple-image system is either a spectroscopic redshift from the CLASH VLT-VIMOS program \citep{balestra:2013}, a Bayesian photometric redshift determined from HST photometry \citep{benitez:2000}, or a value culled from the literature.  Using the method outlined in \citet{merten:2009}, the multiple-image systems are used to infer the location of the critical lines.  The locations of the critical lines are inputs to the reconstruction algorithm.

The weak lensing input takes the form of a shear catalog that lists the coordinates, redshift, and complex ellipticity of background galaxies in the cluster field.  The creation of the HST shear catalog is outlined in Section 3.2 of \citet{merten:2015} and the creation of the Subaru shear catalog is outlined in Section 4 of \citet{umetsu:2014}.  The HST and Subaru catalogs are combined into a single catalog.  Before doing so, the HST complex ellipticity measurements are multiplied by a scale factor to refer them to the effective redshift of the Subaru catalog.  The catalogs are concatenated and the signal-to-noise-weighted mean is computed for sources that appear in both catalogs.

The \texttt{SaWLens} algorithm \citep{merten:2009} is used to perform a nonparametric reconstruction of the lensing potential $\psi(\vect{\theta})$ on an adaptively refined two-dimensional grid from the strong lensing critical lines and the weak lensing shear catalog.  Three different grid sizes are employed: a coarse resolution grid ($25-36$ arcsec pixel), which is applicable to the wide field Subaru weak lensing data, an intermediate resolution grid ($8-13$ arcsec pixel), which is applicable to the HST weak lensing data, and a fine resolution grid ($6-10$ arcsec pixel), which is applicable to the HST strong lensing data.  The lensing potential at each pixel of the grid is estimated by minimizing a $\chi^2$ function that accounts for measurements of the average ellipticity of nearby background galaxies and the location of nearby critical lines.  The assumption of spherical symmetry is not used in this reconstruction, nor are any other prior assumptions about the mass distribution of the cluster.  The convergence of the lens $\kappa(\vect{\theta})$ is then obtained by taking second-order numerical derivatives of the reconstructed lensing potential as prescribed by \Fref{eq:lensing_convergence}.  The \texttt{SaWLens} algorithm has been shown to recover the convergence (or, equivalently, surface mass density) of simulated clusters over a wide range of scales ($50 \ \mbox{kpc} - \mbox{several Mpc}$) with an accuracy of $10\%$ \citep{meneghetti:2010}.

The convergence map is azimuthally binned about the coordinates given in \Fref{tab:spherical_sample}.  The inner boundary is set by the resolution of the highest refinement level of the adaptive grid.  The outer boundary is fixed at the angular scale corresponding to $2 \ \mbox{Mpc} \ h^{-1} \approx 2.85 \ \mbox{Mpc} \ h_{70}^{-1}$.  The radial range defined by these two boundaries is split into 15 bins, with the bin width decreasing as the level of refinement is increased.  

Errors are estimated from 1000 resampled realizations of the $\kappa(\vect{\theta})$ map.  Each realization is created by taking a boot-strap resampling of the shear catalog in the case of weak lensing and a random sampling of the allowed redshift range of the multiple-image systems in the case of strong lensing.  The full reconstruction process and azimuthal binning is carried out on the 1000 realizations.  The set is used to estimate the covariance matrix $\matr{C_{\kappa}}$  of the 15 radial bins.  The convergence profile $\vect{\kappa}$ and associated covariance matrix $\matr{C_{\kappa}}$ then act as inputs to the multiwavelength analysis.

The only difference in the procedure outlined above and that presented in \citet{merten:2015} is that we center the convergence profile on the peak of the X-ray emission rather than the peak of the convergence map.  As a result, we measure a lower convergence in the innermost bin than what is presented in that work.  The choice of center does not have a significant effect on the convergence profile beyond the innermost bin.

\section{Method}
\label{sec:jaco_method}

\subsection{Joint Analysis of Cluster Observations (JACO)}

{
\renewcommand{\arraystretch}{1.20}
\begin{deluxetable*}{ l l l l p{8cm} }
\tablecolumns{5}
\tablecaption{Model parameters and their priors. \label{tab:parameters}}
\tablehead{Parameter & 
           \colhead{Lower Boundary} & 
           \colhead{Upper Boundary} & 
           \colhead{Units}       & 
           Description}
\startdata	
\multicolumn{5}{l}{\textbf{Total Density}} \\[+1mm]
\quad \quad $M\subsc{tot}(0.5 \ \mbox{Mpc})$   & 0.05   & 100.0  & $10^{14} \ \mbox{M}_{\sun}$    & NFW normalization.  Total mass within 0.5 Mpc.  \\
\quad \quad $r_{s}$                 & 0.05   & 25.0   & Mpc                                       & NFW scale radius.  \\[+2mm]
\multicolumn{5}{l}{\textbf{Gas Density}} \\[+1mm]
\quad \quad $M\subsc{gas}(0.5 \ \mbox{Mpc})$   & 0.0001 & 1.0    & $10^{14} \ \mbox{M}_{\sun}$    & Total gas mass within 0.5 Mpc. \\
\quad \quad $r\subsc{gas}$                 & 0.0005 & 2.0    & Mpc                                & Scale radius of the modified $\beta$-model. \\ 
\quad \quad $\beta$                 & 0.30   & 5.0    & \nodata                                   & Power-law slope ($-3\beta$) of the modified $\beta$-model. \\
\quad \quad $r\subsc{gas,\ outer}$                 & 0.20   & 5.0    & Mpc                        & Scale radius of the outer portion of the modified $\beta$-model. \\
\quad \quad $\epsilon$              & 0.20   & 5.0    & \nodata                                   & Power-law slope ($-\epsilon$) of the outer portion of the modified $\beta$-model. \\
\quad \quad $\alpha$                & 0      & 1.5    & \nodata                                   & Power-law slope ($-\alpha$) of the inner portion of the modified $\beta$-model. \\
\quad \quad $[M\subsc{gas,\ core} / M\subsc{gas}](0.5 \ \mbox{Mpc})$ & 0      & 0.50   & \nodata  & Fraction of the total gas mass within 0.5 Mpc that is attributed to the secondary, core $\beta$-model. \\
\quad \quad $r\subsc{gas,\ core}$                 & 0.05   & 50     & kpc                         & Scale radius of the secondary, core $\beta$-model. \\[+2mm]
\multicolumn{5}{l}{\textbf{Nonthermal Pressure Fraction}} \\[+1mm]
\quad \quad $C$                     & 0.00   & 1.825  & \nodata                                   & Normalization of the mean nonthermal pressure fraction profile observed in simulation. \\
\quad \quad $D$                     & 0.00   & 0.50   & \nodata                                   & Normalization of the core nonthermal pressure fraction profile. \\
\quad \quad $E$                     & 0.001  & 0.10   & $r_{200m}$                                & Scale radius of the core nonthermal pressure fraction profile. \\
\quad \quad $\zeta$                 & 0.5    & 3.00   & \nodata                                   & Power law slope ($-\zeta$) of the core nonthermal pressure fraction profile. \\[+2mm]
\multicolumn{5}{l}{\textbf{Nuisance Parameters}} \\[+1mm]
\quad \quad $T\subsc{trunc}$        & 0.00   & 15.0   & keV                                       & Temperature of the ICM at the truncation radius. \\
\quad \quad $Z_{0}$                 & 0.1    & 2.90   & $Z_{\sun}$                                & Metallicity in the center of the cluster. \\
\quad \quad $r_{Z}$                 & 0.005  & 1.00   & Mpc                                       & Metallicity scale radius. \\
\quad \quad $\beta_{Z}$             & 0.00   & 0.80   & \nodata                                   & Metallicity power-law slope ($-3\beta_{Z}$). \\ 
\quad \quad $\bar{T}\subsc{SZ}$     & -1000 & 1000 & $\mu \mbox{K}\subsc{CMB}$               & Mean value of the SZ image. \\
\quad \quad $T\subsc{sbkg}$          & 0.1    & 0.50   & K                                         & Temperature of the soft X-ray background. \\
\quad \quad $A\subsc{sbkg}$          & -0.001 & 0.001  & \nodata                                   & Normalization of the soft X-ray background. \\
\enddata
\tablecomments{Only a subset of these parameters are allowed to float for a given cluster, as determined by the $F$-test decision tree described in \Fref{sec:jaco_model_determination}.  We assume a uniform prior between the lower and upper boundaries.}
\end{deluxetable*}
}

We use the Joint Analysis of Cluster Observations (JACO) software package to fit the model outlined in \Fref{sec:jaco_model} to the X-ray, SZ, and lensing data described in \Fref{sec:jaco_data_description}.  JACO provides a self-consistent framework for modeling and fitting multiwavelength observations of galaxy clusters \citep{mahdavi:2007}.  The general principle underlying JACO is ``forward model fitting''.  The candidate model is projected, convolved, and filtered so that it can be compared to the data directly.  The software is well tested; JACO has been used to examine X-ray and weak lensing scaling relations for a sample of 50 massive galaxy clusters in the Canadian Cluster Comparison Project \citep{mahdavi:2013}.  It has also been used to estimate the hydrostatic mass, gas mass fraction, and ICM temperature from \emph{Chandra} and XMM observations of the CLASH sample \citep{donahue:2014}.  

As part of this work, we have expanded and modified the version of JACO described in \citet{mahdavi:2007,mahdavi:2013} in the following ways.  We have added the ability to fit Bolocam SZ images.  We use the convergence rather than the tangential shear as the lensing observable.  We use a slightly different parameterization for the gas density.  We include nonthermal pressure support in our model.  Finally, although not a change to the underlying JACO package, we include constraints from both weak and strong lensing rather than the weak lensing-only constraints used in previous analyses.

JACO employs a Markov Chain Monte Carlo (MCMC) algorithm to perform Metropolis-Hastings sampling of the joint posterior distribution 
\begin{align}
p(\vect{\theta_{p}} | \vect{S}, \vect{T\subsc{SZ}}, \vect{\kappa}) & \propto \mathpzc{L}\left(\vect{S}, \vect{T\subsc{SZ}}, \vect{\kappa} | \vect{\theta_{p}} \right) \pi({\vect{\theta_{p}}}) \ ,
\end{align}
where $\vect{\theta_{p}}$ is the set of all model parameters, $\mathpzc{L}\left(\vect{\theta_{p}} | \vect{S}, \vect{T\subsc{SZ}}, \vect{\kappa}\right)$ is the likelihood function, and $\pi(\vect{\theta_{p}})$ is the set of prior constraints for the model parameters.  The likelihood function is, up to an overall normalization, given by
\begin{align}
\mathpzc{L}\left(\vect{\theta_{p}} | \vect{S}, \vect{T\subsc{SZ}}, \vect{\kappa}\right) & \propto \exp{\left(-\chi^2\right)},
\end{align}
where
\begin{align}
\label{eq:chisq_total}
\chi^2 & = \chi\subsc{XR}^2 + \chi\subsc{SZ}^2 + \chi\subsc{GL}^2 \ .
\end{align}
That is, we assume that the X-ray, SZ, and lensing measurements are independent, and therefore the total $\chi^2$ is the sum of the $\chi^2$ of the individual datasets.  We now describe how the $\chi^2$ of each dataset is calculated for a candidate model.

For a given set of parameters, JACO generates a set of synthetic X-ray event spectra  $\vect{\widehat{S}}(\vect{\theta_{p}})$ using \Fref{eq:xray_proj} and the input ARF and RMF files.  The cooling function is computed using the MEKAL plasma code.  The model spectra are convolved with the energy-dependent instrument PSF.  The details of how the PSF is calculated for a given set of annular bins can be found in \citet{mahdavi:2007}.  The X-ray contribution to $\chi^2$ is then given by 
\begin{align}
\label{eq:chisq_xr}
\chi\subsc{XR}^2 & =  \displaystyle\sum_{i,j} \frac{(S_{ij} - \widehat{S}_{ij}(\vect{\theta_{p}}))^2}{\sigma_{S_{ij}}^2} \ ,
\end{align}
where the summation runs over the desired annular bins and energy bins.

\begin{figure*}[p]
    \centering
    \includegraphics[width=0.95\textwidth,keepaspectratio]
                    {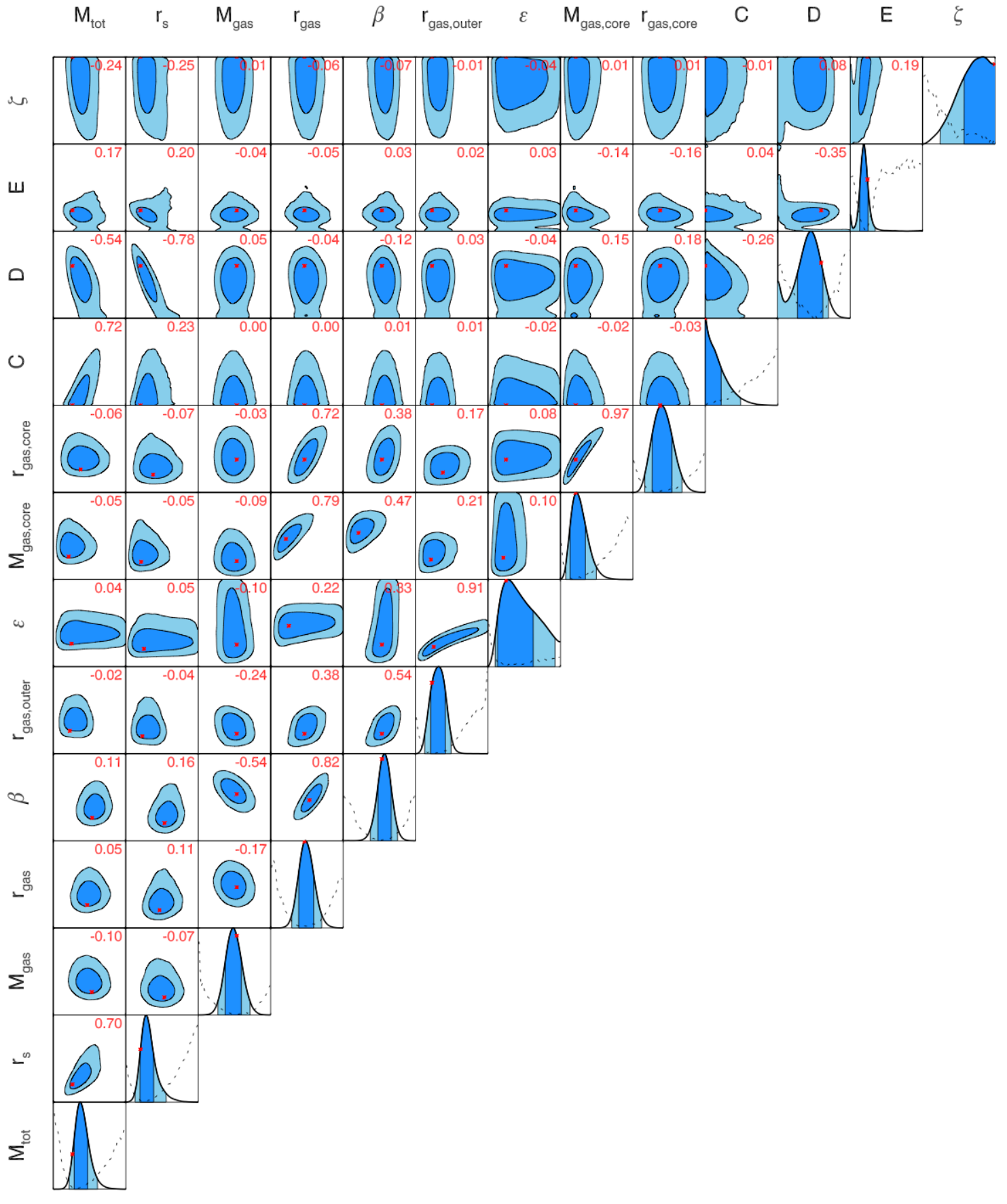}
    \caption{The marginalized two-dimensional joint posterior distributions for the parameters of interest from a fit to the complete multiwavelength dataset for MACS~J1532.8+3021.  Contours denote $68\%$ and $95\%$ credible regions.  Red stars denote the maximum likelihood values.  The red annotation in the upper right corner of each panel is the Pearson correlation coefficient between the two parameters.  The diagonal panels show the marginalized one-dimensional posterior distributions and the dashed black line denotes the negative log-likelihood profile of that parameter.  The purpose of this plot is to illustrate the various parameter degeneracies and the general shape of the posterior distribution.  We have foregone units and tick-marks to simplify the presentation.  Starting from the left column (or bottom row), $M\subsc{tot}$ and $r_{s}$ parameterize the NFW profile for the total density.  The next seven parameters describe the gas density profile: $M\subsc{gas}$, $\beta$, and $r\subsc{gas}$ parameterize intermediate radii, $r\subsc{gas,outer}$ and $\epsilon$ parameterize the steepening of the profile at large radii, and $M\subsc{gas,core}$ and $r\subsc{gas,core}$ parameterize the core.  Finally, $C$, $D$, $E$, and $\zeta$ describe the nonthermal pressure, with $C$ normalizing the bulk term and $D$, $E$, and $\zeta$ parameterizing the small-radius term.}
    \label{fig:joint_posterior_distribution}
\end{figure*}

For a given set of parameters, JACO generates a model SZ image $\vect{\widehat{T}\subsc{SZ}}(\vect{\theta_{p}})$ using Equations~\eqref{eq:sz_proj1}$-$\eqref{eq:sz_proj3}.  Prior to calculating $\chi\subsc{SZ}^2$, it accounts for instrumental effects by simulating the act of observing the model SZ image with Bolocam.  The model image is generated to have a larger size ($25 \ \mbox{arcmin} \times 25 \ \mbox{arcmin}$) and a finer resolution (10 arcsec) than the data to avoid edge effects and sampling effects during convolution.   It is is convolved with a Gaussian kernel with a 60.33 arcsec FWHM in order to account for the instrument PSF (59.17 arcsec FWHM) and pointing uncertainty (5 arcsec RMS).  Afterwards it is rebinned and resized to an identical grid as that of the data.  It is then convolved with the transfer function of the data processing pipeline.  Finally, the parameter $\DCoffset$ is added to the image to represent the unknown mean signal offset.  The SZ contribution to $\chi^2$ is calculated as
\begin{align}
\label{eq:chisq_sz}
\chi\subsc{SZ}^{2} & = \displaystyle\sum_{i} \frac{(T\subsc{SZ,}\,_{i} - \widehat{T}\subsc{SZ,}\,_{i}(\vect{\theta_{p}}))^2}{(C_{T\subsc{SZ}})_{ii}} \ ,
\end{align}
where the summation runs over all pixels with an angular separation $\theta \leq 6.33 \ \mbox{arcmin}$.

Finally, for a given set of parameters, JACO generates a convergence profile $\vect{\widehat{\kappa}}(\vect{\theta_{p}})$ using \Fref{eq:lensing_proj}.  This is compared directly to the convergence profile determined by the \texttt{SaWLens} algorithm.  The lensing contribution to $\chi^2$ is calculated as
\begin{align}
\label{eq:chisq_sw}
\chi\subsc{GL}^2 = \left(\vect{\kappa} - \vect{\widehat{\kappa}}(\vect{\theta_{p}})\right)^{\intercal}  \ \matr{C_{\kappa}}^{-1} \ \left(\vect{\kappa} - \vect{\widehat{\kappa}}(\vect{\theta_{p}})\right) \ ,
\end{align}
which accounts for the nonzero covariance between the radial bins that has been calculated using
the \texttt{SaWLens} bootstraps.

We place a uniform prior on each parameter with the lower and upper boundaries chosen so that the prior is uninformative.  Specifically, the lower and upper boundaries are chosen so that they eliminate regions of parameter space where the likelihood function is already small.  This is not always possible, and in these cases we choose physically reasonable lower and upper boundaries (e.g., the boundaries for the normalization of the nonthermal pressure fraction $C$ are chosen to ensure that $0 \le \mathpzc{F}(r) \le 1$).  The model parameters and their priors are summarized in \Fref{tab:parameters}.  We marginalize  over the nuisance parameters to obtain constraints on the parameters of interest.  \Fref{fig:joint_posterior_distribution} shows an example of the marginalized two-dimensional joint posterior distributions resulting from a JACO fit to the full multiwavelength dataset for MACS~J1532.8+3021.

\subsection{Model Determination}
\label{sec:jaco_model_determination}

The model presented in \Fref{sec:jaco_model} assumes that there is a discrete boundary at which the ICM ends, which we call the truncation radius $r\subsc{trunc}$.  We fix the truncation radius at a distinct physical radius for each cluster that is chosen to be large enough that increasing it further does not have an effect on the model fit.  This is accomplished through the following procedure.  First, we use JACO to fit the NFW model for the total density to the lensing data only.  From these fits, we obtain an estimate of $r_{500c}$.  We then refit the full multiwavelength dataset with the value of $r\subsc{trunc}$ fixed at integer multiples of $r_{500c}$ between 3 and 10.  In all cases, it was found that the resulting constraints on the thermodynamic properties of the ICM converged for values of $r\subsc{trunc} \ge 7 \times r_{500c}$.  We fix the radius at which we truncate the ICM to the physical radius corresponding to $r\subsc{trunc} = 7 \times r_{500c}$ for all further analysis.

The data does not warrant the full complexity of the model presented in \Fref{sec:jaco_model} for any of the clusters in our sample.  We perform a series of $F$-test decision trees in order to determine the maximally restricted model that provides an adequate fit to the data.  The $F$-test is a statistical test that can be used to quantify whether adding additional model parameters results in a significantly better fit to the data.  The test statistic is the fractional increase in the minimum $\chi^2$ that results from restricting the additional parameters divided by the fractional change in the number of degrees of freedom
\begin{align}
F = \frac{\left(\chi\subsc{restricted}^2 - \chi\subsc{unrestricted}^2 \right) / \chi\subsc{unrestricted}^2}{\left(\nu\subsc{restricted} - \nu\subsc{unrestricted} \right) / \nu\subsc{unrestricted}} \ .
\end{align}
The test statistic will follow an $F$-distribution, $F(\nu\subsc{restricted} - \nu\subsc{unrestricted}, \nu\subsc{unrestricted})$, under the null hypothesis that the unrestricted model does not provide a significantly better fit than the restricted model.  We reject the null hypothesis and add the additional model parameters if the probability of observing the measured value of $F$ is less than 0.02.  We apply the $F$-test a total of 48 times in the process of determining the maximally restricted model for all 6 clusters.  The 0.02 cutoff implies that we will add additional model parameters unnecessarily approximately one time.

\begin{deluxetable}{ l c c }
\tablecolumns{3}
\tablecaption{Maximally restricted model for each cluster as determined by the $F$-test decision trees. \label{tab:cluster_model}}
\tablehead{Name & \colhead{Gas}     & \colhead{Nonthermal} \\ 
                & \colhead{Density} & \colhead{Pressure Fraction}}
\startdata
Abell 383           &  G-1b  & F-1a \\
Abell 611           &  G-1a  & F-0 \\
MACS J0429.6-0253   &  G-1a  & F-0 \\
MACS J1311.0-0310   &  G-0   & F-0 \\
MACS J1423.8+2404   &  G-1b  & F-0 \\
MACS J1532.8+3021   &  G-1b  & F-1b
\enddata
\end{deluxetable}

The first $F$-test decision tree is used to determine if the $r^{-\alpha}$ power-law and the second $\beta$-model are necessary to describe the gas density in the cluster core.  We construct the following hierarchy of models ordered by the number of free parameters:
\begin{itemize}
	\item [\textbf{G-0}]  We fix $\alpha = 0$ and $\rho\subsc{gas,c} = 0$. 
	\item [\textbf{G-1a}] We let $\alpha$ float, but fix $\rho\subsc{gas,c} = 0$.
	\item [\textbf{G-1b}] We let $\rho\subsc{gas,c}$ and $r_{c}$ float (recall that $\beta_{c} = 1$), but fix $\alpha = 0$.
	\item [\textbf{G-2}]  We let $\alpha$, $\rho\subsc{gas,c}$, and $r_{c}$ float.
\end{itemize}
We fit all four models to the data.  Since constraints on $\rho\subsc{gas}$ originate from the X-ray and SZ data, we perform this test without the lensing data and assume entirely thermal pressure support.  Since the various models differ only in their treatment of the cluster core, the results of the test are driven almost entirely by the X-ray data.  We examine the two branches of the tree: 0$\rightarrow$1a$\rightarrow$2 and  0$\rightarrow$1b$\rightarrow$2.  We move along each branch, applying the $F$-test at each step, and stop when we either accept the restricted model or reach the end of the branch.  We then compare the stopping points on each branch and choose the model that yields an acceptable fit to the data with the fewest parameters.

After we have settled on a model for the gas density, we carry out a second $F$-test decision tree to determine if a nonthermal pressure component is necessary.  In this case, the hierarchy of models is
\begin{itemize}
	\item [\textbf{F-0}]  We assume completely thermal pressure support by fixing $C=0$ and $D=0$.
	\item [\textbf{F-1a}] We allow for an outer nonthermal pressure component by floating $C$, but fix $D=0$.
	\item [\textbf{F-1b}] We allow for an inner nonthermal pressure component by floating $D$, $E$, and $\zeta$, but fix $C=0$.
	\item [\textbf{F-2}]  We allow for both outer and inner nonthermal pressure components by floating $C$, $D$, $E$, and $\zeta$.
\end{itemize}
We fit all four models to the full multiwavelength dataset and apply the $F$-test decision tree in an identical manner as was carried out for the gas density.  \Fref{tab:cluster_model} lists the maximally restricted model for both the gas density and nonthermal pressure fraction that was chosen for each cluster.  We have compared the constraints on $C$ obtained when fitting model F-1a and model F-2 and find that they are nearly identical.  This suggests that the constraints on $C$ are not driven by the core region of the cluster.

\subsection{SZ Covariance}
\label{sec:sz_covariance}

In order to determine the effect that the small inter-pixel correlations in the SZ image have on our results, we have carried out the following simulation for the galaxy cluster Abell~611.  We take the best-fit maximally restricted model and generate 100 model-plus-noise realizations.  In the case of the X-ray data, this is accomplished by perturbing the model prediction for each X-ray spectral bin $\hat{S}_{ij}(\vect{\theta}_{p})$ by a random draw from a Gaussian with mean equal to zero and standard deviation equal to $\sigma_{S_{ij}}$.  In the case of the lensing data, this is accomplished by perturbing the model prediction for the convergence profile $\vect{\hat{\kappa}}(\vect{\theta}_{p})$ by a random draw from a multivariate Gaussian distribution with mean equal to zero and covariance equal to $\matr{C_{\kappa}}$.  Finally, in the case of the SZ data, this is accomplished by adding a random noise realization to the model prediction for the SZ image $\vect{\hat{T}}\subsc{SZ}(\vect{\theta}_{p})$.  The SZ noise realizations are described in \Fref{sec:bolocam_sz}; recall that they contain the inter-pixel correlations that this simulation aims to understand.  For each of the 100 model-plus-noise realizations, we repeat the full JACO fit.  We then compare the resulting distribution of best-fit parameter values to the marginalized posterior distribution obtained from the original fit to the data (which assumes a diagonal covariance matrix for the SZ data).  We find no significant bias in the center of the distribution for the parameters of interest.  More specifically, for each parameter of interest, the center of the distribution of best-fit values obtained from fitting the 100 model-plus-noise realizations, which contain the inter-pixel SZ correlations, differs from the center of the marginalized posterior distribution of the original fit to the data, which assumes a diagonal SZ covariance matrix, at roughly $10\%$ of the width of the marginalized posterior distribution.  This is consistent with our uncertainty on the quantity due to the fact that we have a sample size of 100.  Similarly, we find no significant change in the width of the distribution for the parameters of interest.  The widths estimated with and without SZ correlations differ at roughly the $10\%$ level, again consistent with how well we can measure this quantity as estimated by bootstrap resampling the 100 samples.  Note that the choice of 100 samples was a balance between computation time and resulting sensitivity.  We have assumed that the conclusions drawn from this simulation generalize to the other clusters in our sample, and thus we assume a diagonal SZ covariance matrix for the results presented in the following section.

\section{Results}
\label{sec:jaco_results}

\begin{deluxetable*}{ l c c c c c c c c }
    \vspace{2.5cm}
\tablecolumns{9}
\tablecaption{Quality of fit to different combinations of datasets. \label{tab:jaco_fit_quality}}
\tablehead{Name & 
           \colhead{$\chi^2_{\mbox{\tiny XR}}$} & 
           \colhead{$\chi^2_{\mbox{\tiny SZ}}$} & 
           \colhead{$\chi^2_{\mbox{\tiny GL}}$} & 
           \colhead{$\chi^2$}                   & 
           \colhead{$N$}                        & 
           \colhead{$N_{\mbox{\tiny param}}$}   & 
           \colhead{$\nu$}                      & 
           \colhead{$\mbox{PTE}$}}
\startdata
\textbf{Abell 383} &  &  &  &  &  &  &  & \\
    \quad GL                      & \nodata & \nodata & 2.0 & 2.0 & 15 & 2 & 13 & 1.00 \\ 
    \quad XR                      & 1636.4 & \nodata & \nodata & 1636.4 & 1477 & 16 & 1461 & 0.00086 \\ 
    \quad XR+SZ                   & 1637.8 & 1203.3 & \nodata & 2841.1 & 2601 & 16 & 2585 & 0.00027 \\ 
    \quad XR+SZ+GL                & 1636.7 & 1201.9 & 6.9 & 2845.5 & 2616 & 17 & 2599 & 0.00044 \\ 
    \quad XR+SZ+GL (Nonthermal)   & 1636.7 & 1201.9 & 6.9 & 2845.5 & 2616 & 17 & 2599 & 0.00044 \\ [+2mm]
\textbf{Abell 611} &  &  &  &  &  &  &  & \\
    \quad GL                      & \nodata & \nodata & 4.2 & 4.2 & 15 & 2 & 13 & 0.99 \\ 
    \quad XR                      & 1015.0 & \nodata & \nodata & 1015.0 & 1037 & 14 & 1023 & 0.56 \\ 
    \quad XR+SZ                   & 1016.1 & 1134.9 & \nodata & 2150.9 & 2161 & 14 & 2147 & 0.47 \\ 
    \quad XR+SZ+GL                & 1016.3 & 1135.6 & 7.9 & 2159.8 & 2176 & 14 & 2162 & 0.51 \\ 
    \quad XR+SZ+GL (Nonthermal)   & 1016.5 & 1135.3 & 8.0 & 2159.7 & 2176 & 15 & 2161 & 0.50 \\ [+2mm]
\textbf{MACS J0429.6-0253} &  &  &  &  &  &  &  & \\
    \quad GL                      & \nodata & \nodata & 2.9 & 2.9 & 15 & 2 & 13 & 1.00 \\ 
    \quad XR                      & 246.7 & \nodata & \nodata & 246.7 & 258 & 14 & 244 & 0.44 \\ 
    \quad XR+SZ                   & 248.7 & 1200.2 & \nodata & 1448.9 & 1382 & 14 & 1368 & 0.063 \\ 
    \quad XR+SZ+GL                & 249.2 & 1200.0 & 5.4 & 1454.6 & 1397 & 14 & 1383 & 0.088 \\ 
    \quad XR+SZ+GL (Nonthermal)   & 249.2 & 1200.0 & 5.4 & 1454.6 & 1397 & 14 & 1382 & 0.085 \\ [+2mm]
\textbf{MACS J1311.0-0310} &  &  &  &  &  &  &  & \\
    \quad GL                      & \nodata & \nodata & 3.8 & 3.8 & 15 & 2 & 13 & 0.99 \\ 
    \quad XR                      & 295.8 & \nodata & \nodata & 295.8 & 337 & 13 & 324 & 0.87 \\ 
    \quad XR+SZ                   & 297.2 & 1143.3 & \nodata & 1440.5 & 1461 & 13 & 1448 & 0.55 \\ 
    \quad XR+SZ+GL                & 297.1 & 1143.4 & 3.9 & 1444.4 & 1476 & 13 & 1463 & 0.63 \\ 
    \quad XR+SZ+GL (Nonthermal)   & 297.1 & 1143.4 & 3.9 & 1444.4 & 1476 & 13 & 1462 & 0.62 \\ [+2mm]
\textbf{MACS J1423.8+2404} &  &  &  &  &  &  &  & \\
    \quad GL                      & \nodata & \nodata & 6.4 & 6.4 & 15 & 2 & 13 & 0.93 \\ 
    \quad XR                      & 820.9 & \nodata & \nodata & 820.9 & 909 & 15 & 894 & 0.96 \\ 
    \quad XR+SZ                   & 824.5 & 1076.0 & \nodata & 1900.5 & 2033 & 15 & 2018 & 0.97 \\ 
    \quad XR+SZ+GL                & 823.2 & 1076.5 & 7.2 & 1907.0 & 2048 & 15 & 2033 & 0.98 \\ 
    \quad XR+SZ+GL (Nonthermal)   & 823.4 & 1075.9 & 7.4 & 1906.7 & 2048 & 16 & 2032 & 0.98 \\ [+2mm]
\textbf{MACS J1532.8+3021} &  &  &  &  &  &  &  & \\
    \quad GL                      & \nodata & \nodata & 5.2 & 5.2 & 15 & 2 & 13 & 0.97 \\ 
    \quad XR                      & 2708.1 & \nodata & \nodata & 2708.1 & 2808 & 15 & 2793 & 0.87 \\ 
    \quad XR+SZ                   & 2719.9 & 1249.0 & \nodata & 3968.9 & 3932 & 15 & 3917 & 0.28 \\ 
    \quad XR+SZ+GL                & 2704.8 & 1245.1 & 17.6 & 3967.5 & 3947 & 18 & 3929 & 0.33 \\ 
    \quad XR+SZ+GL (Nonthermal)   & 2704.8 & 1245.1 & 17.6 & 3967.5 & 3947 & 19 & 3928 & 0.33
\enddata
\tablecomments{For each galaxy cluster in the sample we tabulate the quality of fit to the lensing data only (GL), X-ray data only (XR), joint X-ray and SZ data (XR+SZ), full multiwavelength dataset using the maximally restricted model (XR+SZ+GL), and full multiwavelength dataset using the maximally restricted model including an outer nonthermal pressure component (XR+SZ+GL (Nonthermal)).  The columns denote, from left to right: $\chi^2$ for the X-ray data (see \Fref{eq:chisq_xr}), $\chi^2$ for the SZ data (see \Fref{eq:chisq_sz}), $\chi^2$ for the lensing data (see \Fref{eq:chisq_sw}), total $\chi^2$ (see \Fref{eq:chisq_total}), number of data points $N$, number of parameters $N\subsc{param}$, number of degrees of freedom $\nu = N - N\subsc{param}$, and the probability to exceed (PTE) the total $\chi^2$ based on the $\chi^2(\nu)$ probability density function.  In the case of the GL-only fits, the low $\chi^2$ values are driven primarily by the data at large radius, where the constraining power is relatively poor (see \citet{merten:2015} for additional details).  \vspace{2.5cm}}
\end{deluxetable*}

In order to investigate the interplay between the various datasets, we fit lensing only (GL), X-ray only (XR), joint X-ray and SZ (XR+SZ), and the full dataset (XR+SZ+GL).  We do not perform an SZ only fit because the SZ data alone is not sufficient to fully constrain the thermodynamic properties of the ICM.  When we fit the full dataset, we use the maximally restricted model determined in \Fref{sec:jaco_model_determination} for each cluster.  When we fit subsets of the full dataset we use restricted versions of this model.  In the case of GL, the model reduces to an NFW density profile fully described by two parameters.  In the case of XR and XR+SZ, we assume entirely thermal pressure support (by fixing $C=0$ and $D=0$) because our ability to constrain the nonthermal pressure component relies on a comparison of the lensing and X-ray/SZ data.  We note that the GL fits use data that are identical to those used by \citet{merten:2015}, other than the choice of cluster center, and our derived parameters from the GL fits are fully consistent with those derived by \citet{merten:2015}. Furthermore, the XR fits use data that are identical to those used by \citet{donahue:2014}, other than the addition of one more annulus at large radius, and the derived parameters from our XR fits are consistent with those derived in \citet{donahue:2014}.

For each fit, we first employ a Levenberg--Marquardt (LM) minimization algorithm to search for the global maximum of the likelihood function.  We then run 8 MCMC chains in parallel all starting from the best-fit parameter values determined by the LM algorithm.  Each chain is run for $22,500 \times N\subsc{param}$ total iterations.  The first $10\%$ of the iterations are discarded as burn-in and the chains are concatenated.  This yields 2--3 million draws from the joint posterior distribution.  The acceptance rate of the MCMC algorithm is close to optimal with approximately $25\%$ of the proposed steps accepted \citep{roberts:2001}.  However, the chains have significant serial correlation; we observe an exponential decay in the autocorrelation function with an $e$-folding time $\tau \sim 1000$ iterations.  We thin the chains by $\tau$ when calculating statistics, which results in an effective sample size of $2,000$--$3,000$.  We apply the Geweke diagnostic \citep{geweke:1992}, Heidelberger-Welch diagnostic \citep{heidelberger:1983,heidelberger:1981}, and Raftery-Lewis diagnostic \citep{raftery:1992} to the individual parameter chains to confirm that they have converged at an acceptable level.

The minimum $\chi^2$ for each fit is presented in \Fref{tab:jaco_fit_quality} along with the number of model parameters, number of degrees of freedom, and the probability to exceed (PTE).  All of the clusters have an acceptable quality of fit for all of the data combinations, with the exception of Abell 383.  There is modest tension between the X-ray and SZ data for MACS J0429.6-0253 and MACS J1532.8+3021, which is evident in the decrease in PTE when including the SZ data (XR $\rightarrow$ XR+SZ).  We address this tension in the subsections below where we discuss each cluster individually.  The best-fit models corresponding to the XR+SZ+GL rows are compared to the data in Appendix~\ref{sec:best_fit_model_figs} (Figures~\ref{fig:radial_jaco_fit1}--\ref{fig:radial_jaco_fit5}).

\begin{figure*}[p]
    \centering
    \includegraphics[width=0.8\textwidth,keepaspectratio]{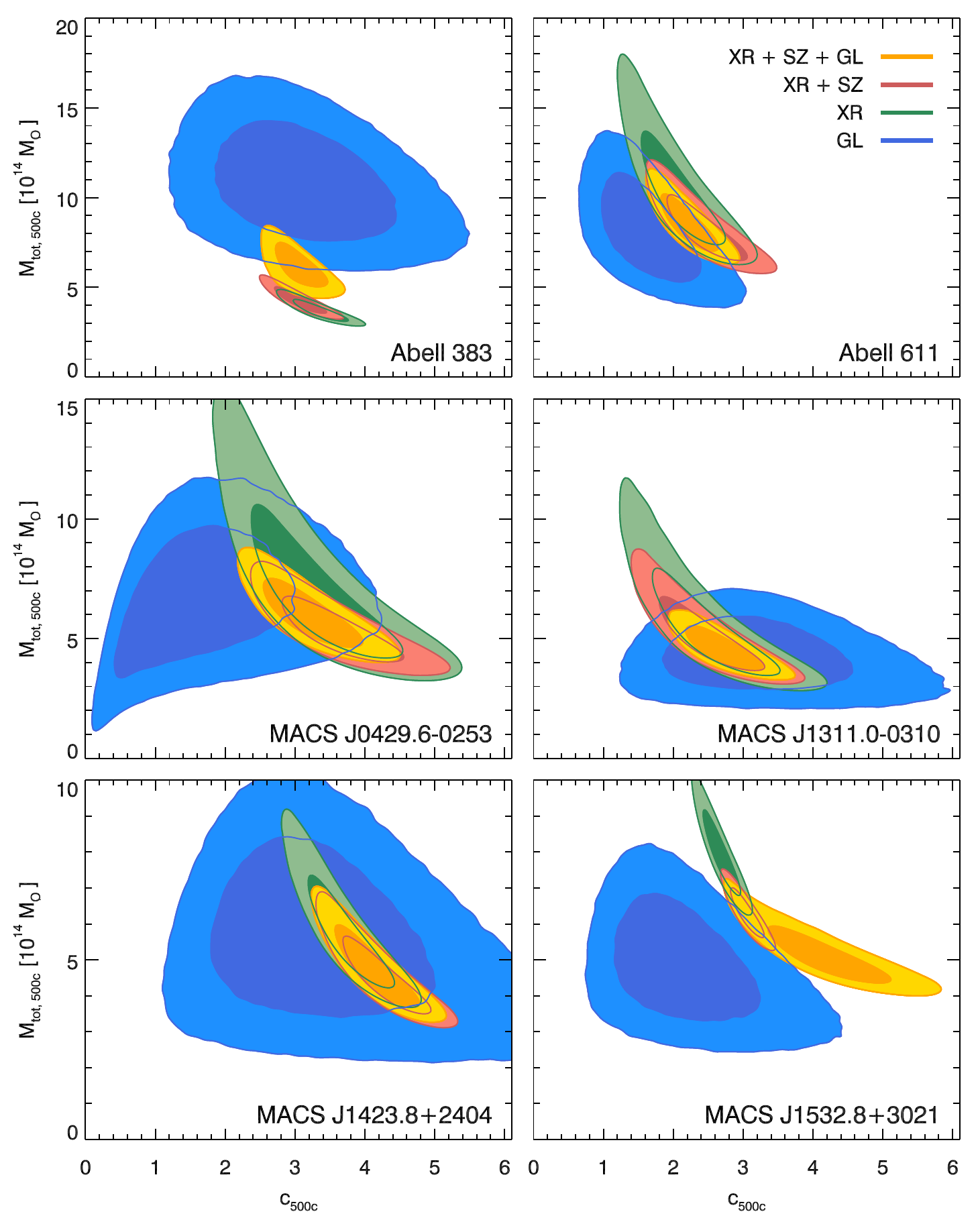}
    \caption{Constraints on the concentration and total mass at $r_{500c}$ for the six galaxy clusters in our sample.  Contours denote $68\%$ and $95\%$ credible regions.  The colors denote fits to different combinations of datasets.  Blue denotes a fit to the lensing data only (GL), green the X-ray data only (XR), red the X-ray and SZ data (XR+SZ), and gold the full multiwavelength dataset using the maximally restricted model (XR+SZ+GL).  Note that the range of the y-axis is different for each row.  In the case of MACS J1532.8+3021, the model employed in the XR+SZ+GL fit includes an inner nonthermal pressure component that was omitted from the other three analyses (because it cannot be constrained without the full multiwavelength dataset) and results in the seemingly conflicting constraints on the concentration.}
    \label{fig:c500_M500}
\end{figure*}

\begin{turnpage}

\begin{deluxetable*}{ l c c c c c c c c c c }
\tablecolumns{11}
\tablecaption{Constraints on the overdensity radius, concentration, total mass, and gas mass fraction at several overdensities. Some values are missing either because they are unconstrained (such as $f_{\mbox{\footnotesize gas}}$ for the GL fit) or beyond the maximum radius that can be reliably constrained by the X-ray data (see Section~\ref{sec:jaco_results}). \label{tab:cM_constraints}}
\tablehead{
Name & \colhead{$r_{2500c}$} & \colhead{$M_{\mbox{\footnotesize tot,} 2500c}$} & \colhead{$f_{\mbox{\footnotesize gas,}2500c}$} & \colhead{$r_{500c}$} & \colhead{$M_{\mbox{\footnotesize tot,} 500c}$} & \colhead{$f_{\mbox{\footnotesize gas,}500c}$} & \colhead{$r_{200c}$} & \colhead{$c_{200c}$} & \colhead{$M_{\mbox{\footnotesize tot,} 200c}$} & \colhead{$f_{\mbox{\footnotesize gas,}200c}$}\\
 & \colhead{$[\mbox{kpc} \ h_{70}^{-1}]$} & \colhead{$[10^{14} \ \mbox{M}_{\odot} \ h_{70}^{-1}]$} & \colhead{} & \colhead{$[\mbox{kpc} \ h_{70}^{-1}]$} & \colhead{$[10^{14} \ \mbox{M}_{\odot} \ h_{70}^{-1}]$} & \colhead{} & \colhead{$[\mbox{kpc} \ h_{70}^{-1}]$} & \colhead{} & \colhead{$[10^{14} \ \mbox{M}_{\odot} \ h_{70}^{-1}]$} & \colhead{}}
\startdata
\textbf{Abell 383}\tablenotemark{a} &  &  &  &  &  &  &  &  &  & \\
\quad GL                      & $645 \pm 45$ & $4.54 \pm 0.95$ &      \nodata      & $1460 \pm 100$ & $10.5 \pm 2.2$ &      \nodata      & $2220 \pm 170$ & $4.6 \pm 1.2$ & $14.8 \pm 3.4$ &      \nodata      \\
\quad XR                      & $460 \pm 10$ & $1.66 \pm 0.13$ & $0.098 \pm 0.004$ & $1020 \pm 35$ & $3.6 \pm 0.4$ & $0.117 \pm 0.011$ &      \nodata      &      \nodata      &      \nodata      &      \nodata      \\
\quad XR+SZ                   & $475 \pm 10$ & $1.83 \pm 0.14$ & $0.093 \pm 0.004$ & $1075 \pm 40$ & $4.2 \pm 0.5$ & $0.104 \pm 0.011$ &      \nodata      &      \nodata      &      \nodata      &      \nodata      \\
\quad XR+SZ+GL                & $535 \pm 20$ & $2.64 \pm 0.27$ & $0.075 \pm 0.005$ & $1210 \pm 50$ & $6.0 \pm 0.8$ & $0.800 \pm 0.010$ & $1840 \pm 85$ & $4.6 \pm 0.3$ & $8.5 \pm 1.2$ &      \nodata      \\
\quad XR+SZ+GL (Nonthermal)   & $535 \pm 20$ & $2.64 \pm 0.27$ & $0.075 \pm 0.005$ & $1210 \pm 50$ & $6.0 \pm 0.8$ & $0.800 \pm 0.010$ & $1840 \pm 85$ & $4.6 \pm 0.3$ & $8.5 \pm 1.2$ &      \nodata      \\[+2mm]
\textbf{Abell 611} &  &  &  &  &  &  &  &  &  & \\
\quad GL                      & $495 \pm 35$ & $2.30 \pm 0.48$ &      \nodata      & $1285 \pm 105$ & $7.9 \pm 2.0$ &      \nodata      & $2035 \pm 200$ & $2.5 \pm 0.7$ & $12.6 \pm 3.8$ &      \nodata      \\
\quad XR                      & $570 \pm 20$ & $3.50 \pm 0.38$ & $0.083 \pm 0.005$ & $1380 \pm 95$ & $9.9 \pm 2.0$ & $0.094 \pm 0.011$ &      \nodata      &      \nodata      &      \nodata      &      \nodata      \\
\quad XR+SZ                   & $545 \pm 15$ & $3.09 \pm 0.24$ & $0.089 \pm 0.004$ & $1280 \pm 60$ & $8.0 \pm 1.2$ & $0.107 \pm 0.009$ &      \nodata      &      \nodata      &      \nodata      &      \nodata      \\
\quad XR+SZ+GL                & $545 \pm 15$ & $3.07 \pm 0.21$ & $0.089 \pm 0.003$ & $1305 \pm 55$ & $8.4 \pm 1.0$ & $0.104 \pm 0.008$ & $2025 \pm 95$ & $3.4 \pm 0.4$ & $12.5 \pm 1.8$ &      \nodata      \\
\quad XR+SZ+GL (Nonthermal)   & $550 \pm 15$ & $3.14 \pm 0.27$ & $0.088 \pm 0.004$ & $1315 \pm 60$ & $8.7 \pm 1.2$ & $0.102 \pm 0.008$ & $2045 \pm 105$ & $3.4 \pm 0.4$ & $12.9 \pm 2.0$ &      \nodata      \\[+2mm]
\textbf{MACS J0429.6-0253} &  &  &  &  &  &  &  &  &  & \\
\quad GL                      & $470 \pm 70$ & $2.04 \pm 0.97$ &      \nodata      & $1160 \pm 110$ & $6.6 \pm 1.9$ &      \nodata      & $1840 \pm 165$ & $2.6 \pm 1.2$ & $10.6 \pm 2.9$ &      \nodata      \\
\quad XR                      & $515 \pm 35$ & $2.95 \pm 0.63$ & $0.095 \pm 0.010$ & $1145 \pm 115$ & $6.4 \pm 2.0$ & $0.116 \pm 0.027$ &      \nodata      &      \nodata      &      \nodata      &      \nodata      \\
\quad XR+SZ                   & $485 \pm 20$ & $2.43 \pm 0.28$ & $0.103 \pm 0.007$ & $1060 \pm 60$ & $5.2 \pm 0.9$ & $0.143 \pm 0.016$ &      \nodata      &      \nodata      &      \nodata      &      \nodata      \\
\quad XR+SZ+GL                & $495 \pm 15$ & $2.61 \pm 0.27$ & $0.099 \pm 0.006$ & $1110 \pm 55$ & $5.9 \pm 0.9$ & $0.132 \pm 0.013$ & $1680 \pm 100$ & $4.8 \pm 0.6$ & $8.2 \pm 1.4$ &      \nodata      \\
\quad XR+SZ+GL (Nonthermal)   & $510 \pm 25$ & $2.90 \pm 0.41$ & $0.093 \pm 0.007$ & $1150 \pm 65$ & $6.5 \pm 1.1$ & $0.125 \pm 0.014$ & $1735 \pm 110$ & $5.0 \pm 0.7$ & $8.9 \pm 1.7$ &      \nodata      \\[+2mm]
\textbf{MACS J1311.0-0310} &  &  &  &  &  &  &  &  &  & \\
\quad GL                      & $425 \pm 35$ & $1.81 \pm 0.47$ &      \nodata      & $960 \pm 75$ & $4.2 \pm 1.0$ &      \nodata      & $1455 \pm 120$ & $4.7 \pm 1.3$ & $5.8 \pm 1.5$ &      \nodata      \\
\quad XR                      & $435 \pm 25$ & $2.00 \pm 0.31$ & $0.099 \pm 0.009$ & $1020 \pm 95$ & $5.0 \pm 1.4$ & $0.107 \pm 0.021$ &      \nodata      &      \nodata      &      \nodata      &      \nodata      \\
\quad XR+SZ                   & $430 \pm 15$ & $1.91 \pm 0.20$ & $0.102 \pm 0.007$ & $1005 \pm 65$ & $4.9 \pm 1.0$ & $0.112 \pm 0.015$ &      \nodata      &      \nodata      &      \nodata      &      \nodata      \\
\quad XR+SZ+GL                & $425 \pm 10$ & $1.84 \pm 0.15$ & $0.104 \pm 0.005$ & $980 \pm 40$ & $4.5 \pm 0.6$ & $0.119 \pm 0.011$ & $1500 \pm 75$ & $4.1 \pm 0.5$ & $6.5 \pm 1.0$ &      \nodata      \\
\quad XR+SZ+GL (Nonthermal)   & $435 \pm 20$ & $2.00 \pm 0.25$ & $0.099 \pm 0.007$ & $1005 \pm 50$ & $4.9 \pm 0.7$ & $0.113 \pm 0.011$ & $1540 \pm 80$ & $4.2 \pm 0.5$ & $7.0 \pm 1.1$ &      \nodata      \\[+2mm]
\textbf{MACS J1423.8+2404} &  &  &  &  &  &  &  &  &  & \\
\quad GL                      & $455 \pm 45$ & $2.40 \pm 0.73$ &      \nodata      & $1025 \pm 100$ & $5.4 \pm 1.6$ &      \nodata      & $1560 \pm 165$ & $4.8 \pm 1.5$ & $7.6 \pm 2.4$ &      \nodata      \\
\quad XR                      & $470 \pm 20$ & $2.63 \pm 0.37$ & $0.100 \pm 0.008$ &      \nodata      &      \nodata      &      \nodata      &      \nodata      &      \nodata      &      \nodata      &      \nodata      \\
\quad XR+SZ                   & $440 \pm 20$ & $2.22 \pm 0.27$ & $0.110 \pm 0.008$ &      \nodata      &      \nodata      &      \nodata      &      \nodata      &      \nodata      &      \nodata      &      \nodata      \\
\quad XR+SZ+GL                & $450 \pm 20$ & $2.31 \pm 0.28$ & $0.108 \pm 0.008$ & $965 \pm 50$ & $4.6 \pm 0.7$ &      \nodata      & $1445 \pm 80$ & $6.1 \pm 0.5$ & $6.2 \pm 1.1$ &      \nodata      \\
\quad XR+SZ+GL (Nonthermal)   & $470 \pm 25$ & $2.69 \pm 0.42$ & $0.098 \pm 0.009$ & $1020 \pm 60$ & $5.4 \pm 1.0$ &      \nodata      & $1525 \pm 95$ & $6.2 \pm 0.5$ & $7.2 \pm 1.4$ &      \nodata      \\[+2mm]
\textbf{MACS J1532.8+3021} &  &  &  &  &  &  &  &  &  & \\
\quad GL                      & $435 \pm 35$ & $1.64 \pm 0.40$ &      \nodata      & $1060 \pm 85$ & $4.7 \pm 1.2$ &      \nodata      & $1655 \pm 150$ & $3.1 \pm 1.0$ & $7.2 \pm 2.0$ &      \nodata      \\
\quad XR                      & $540 \pm 10$ & $3.20 \pm 0.21$ & $0.109 \pm 0.004$ & $1245 \pm 40$ & $7.9 \pm 0.8$ & $0.109 \pm 0.008$ & $1905 \pm 70$ & $4.1 \pm 0.2$ & $11.3 \pm 1.2$ & $0.103 \pm 0.010$ \\
\quad XR+SZ                   & $510 \pm 10$ & $2.72 \pm 0.13$ & $0.120 \pm 0.004$ & $1155 \pm 30$ & $6.2 \pm 0.5$ & $0.128 \pm 0.007$ & $1755 \pm 45$ & $4.6 \pm 0.2$ & $8.8 \pm 0.7$ & $0.129 \pm 0.009$ \\
\quad XR+SZ+GL                & $500 \pm 10$ & $2.51 \pm 0.14$ & $0.126 \pm 0.004$ & $1070 \pm 35$ & $5.0 \pm 0.5$ & $0.150 \pm 0.011$ & $1605 \pm 65$ & $6.2 \pm 0.8$ & $6.7 \pm 0.8$ & $0.162 \pm 0.016$ \\
\quad XR+SZ+GL (Nonthermal)   & $505 \pm 15$ & $2.62 \pm 0.21$ & $0.123 \pm 0.006$ & $1090 \pm 45$ & $5.3 \pm 0.7$ & $0.146 \pm 0.013$ & $1630 \pm 80$ & $6.2 \pm 0.8$ & $7.0 \pm 1.0$ & $0.155 \pm 0.018$ \\[+2mm]
\enddata
\tablenotetext{a}{Abell 383 is not adequately described by the spherical model; use caution when interpreting the results for this cluster.}
\end{deluxetable*}
\end{turnpage}

We present the resulting constraints on the total mass $M\subsc{tot}$, concentration $c$, and gas mass fraction $f\subsc{gas}(r) = M\subsc{gas}(r) / M\subsc{tot}(r)$ at several overdensity radii in \Fref{tab:cM_constraints}.  The quoted value and error correspond to the center and half of the span of the smallest $68\%$ credible region determined from the marginalized posterior distribution for that parameter.  We also plot the two-dimensional constraints on $M\subsc{tot,500c}$--$c_{500c}$ in \Fref{fig:c500_M500}.

As mentioned in \Fref{sec:jaco_model_determination}, Abell 383 is the only cluster that requires an outer nonthermal pressure component based on our $F$-test decision tree.  For this cluster, the total mass inferred from the GL analysis is 2--3 times larger than that inferred from the XR or XR+SZ analysis.  This forces the nonthermal pressure fraction to very large values when performing the XR+SZ+GL analysis, and even that does not resolve the discrepancy, as evidenced by the poor quality of fit.  We do not believe that a spherically symmetric model is a reasonable approximation for Abell 383, for reasons that will be outlined in \Fref{sec:abell_383}.  Both nonthermal pressure support and an elongation of the cluster along the line-of-sight direction will elevate the lensing inferred mass compared to the X-ray/SZ inferred mass.  Hence, if the cluster is elongated along the line-of-sight direction, the nonthermal pressure fraction inferred from a spherical fit will be overestimated.  We do not include Abell 383 in our analysis of the nonthermal pressure support for this reason and stress caution in interpreting the resulting mass estimates.

\begin{figure}[t]
    \centering
    \includegraphics[width=\linewidth,keepaspectratio]{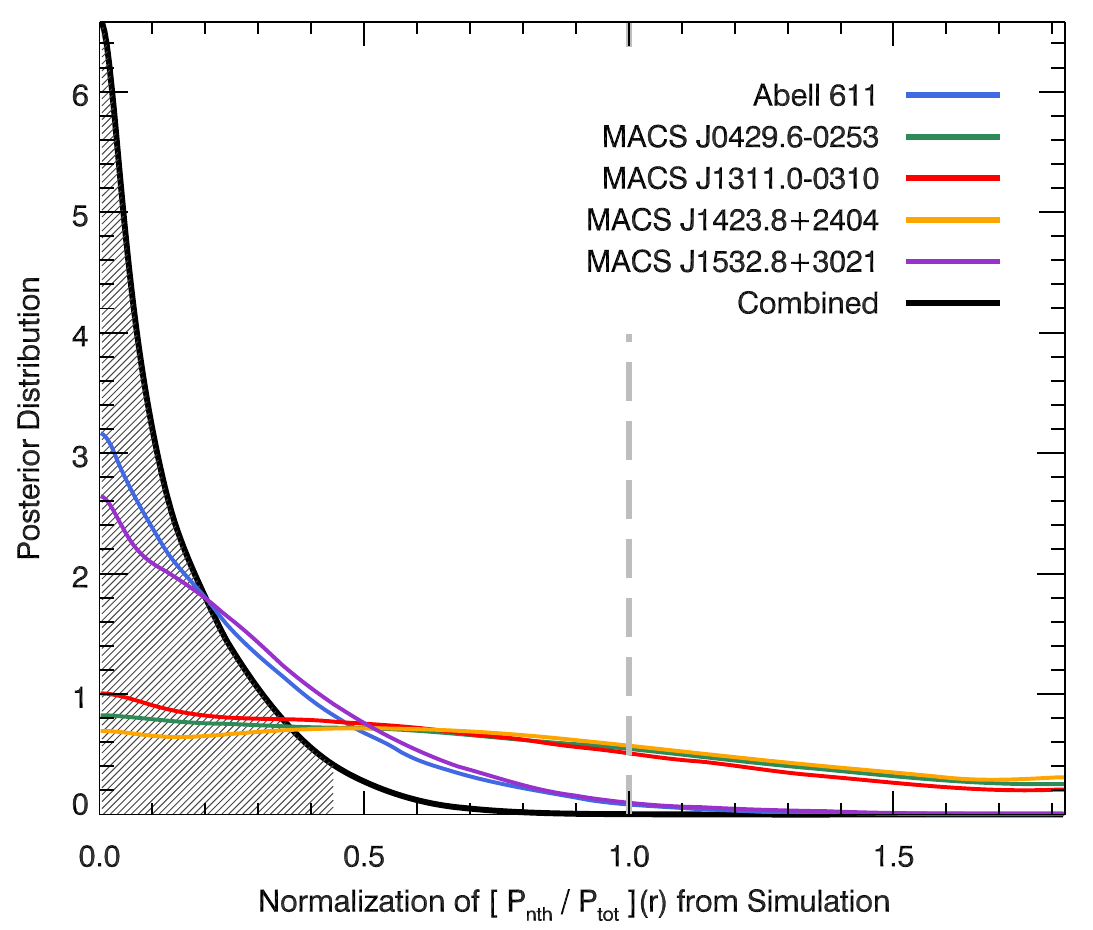}
    \caption{Posterior distribution of the normalization $C$ of the best-fit nonthermal pressure fraction profile from \citet{nelson:2014}.  The different colors denote the different galaxy clusters in the spherical sample (excluding Abell 383).  Black denotes the combined posterior distribution obtained by multiplying the posterior distributions from the individual clusters.  The shading denotes the $95\%$ credible region determined from the combined posterior distribution.  The dashed gray line at $C=1.0$ corresponds to the mean value observed in simulation.} 
    \label{fig:posterior_nonthermal_pressure}
\end{figure}

We use the other five clusters to test for the nonthermal pressure support predicted by simulations.  We perform a second fit to the full multiwavelength dataset allowing the normalization $C$ of the nonthermal pressure fraction profile calibrated from simulation to vary.  This fit is labeled ``XR+SZ+GL (Nonthermal)'' in \Fref{tab:jaco_fit_quality} and \Fref{tab:cM_constraints}.   Note that a uniform prior $U(0, \ 1.825)$ is placed on $C$.  The lower bound $C=0$ corresponds to entirely thermal pressure support at all radii.  The upper bound $C=1.825$ corresponds to zero thermal pressure support at the cluster outskirts ($r \gtrsim r_{200m}$).  The marginalized posterior distribution for $C$ is shown in \Fref{fig:posterior_nonthermal_pressure} for each of the five clusters.  We find that MACS J0429.6-0253, MACS J1311.0-0310, and MACS J1423.8+2404 have fairly flat posterior distributions, although there is a preference for $C$ less than $1.0$ over $C$ greater than $1.0$.  Abell 611 and MACSJ1532.8+3021 have higher quality X-ray data and as a result are able to place meaningful upper bounds on the nonthermal pressure fraction.  Since constraints from the individual clusters are consistent with a common value of $C$, we multiply the individual posterior distributions together to obtain a combined constraint.  The resulting $95\%$ credible interval on the normalization $C$ is $(0, \ 0.43)$.  Hence, the universal nonthermal pressure fraction profile observed in simulations ($C=1.0$) is an extremely unlikely description of this sample of five clusters.  We also derive the combined constraint on the nonthermal pressure fraction $\mathpzc{F}(r)$ at several over-density radii $r=[r_{2500c}, \ r_{500c}, \ r_{200c}, \ r_{200m}]$ using the same procedure.  These are presented in \Fref{tab:nonthermal_pressure_fraction}.

\begin{deluxetable}{ l c c c }
\tablecolumns{4}
\tablecaption{Upper bound on the nonthermal pressure fraction at several overdensity radii. \label{tab:nonthermal_pressure_fraction}}
\tablehead{Parameter &
           \colhead{$N_{\mbox{\tiny cluster}}$\,\tablenotemark{a}} & 
           \colhead{$95\%$ Upper Bound} &
           \colhead{Expectation from Simulation\,\tablenotemark{b}}}
\startdata
$C$                      & 5                        & 0.44                       & 1.00                       \\ 
$\mathpzc{F}(r_{2500c})$ & 5                        & 0.06                       & 0.15                       \\ 
$\mathpzc{F}(r_{500c})$  & 4                        & 0.11                       & 0.26                       \\ 
$\mathpzc{F}(r_{200c})$  & 1                        & 0.29                       & 0.35                       \\ 
$\mathpzc{F}(r_{200m})$  & 1                        & 0.35                       & 0.43                       
\enddata
\tablenotetext{a}{Number of galaxy clusters used to construct the $95\%$ upper bound.}
\tablenotetext{b}{Median value from the simulation of \citet{nelson:2014} for clusters with the mass/redshift as those used to construct the upper bound.}
\end{deluxetable}

While the GL and SZ data are quite uniform over the sample, the radii over which we have X-ray constraints varies significantly from cluster to cluster, depending on the cluster redshift and the total integration time achieved by \emph{Chandra}.  The X-ray data is necessary to constrain the gas density and fully characterize the thermodynamic state of the ICM.  In order to determine the maximum radius where our model provides reliable results, we perform the following test using the two clusters with the highest quality X-ray data, Abell 611 and MACS J1532.8+3021.  We repeat the XR+SZ+GL (Nonthermal) fit multiple times, each time discarding the outermost X-ray annulus.  We compare the thermodynamic profiles obtained from fits to the reduced X-ray datasets to those obtained from the fits to the full X-ray dataset.  Specifically, we examine the total density, gas density, temperature, entropy, pressure, and nonthermal pressure fraction as a function of the ratio of the outer radius of the reduced dataset to the outer radius of the full dataset.  In examining these fits, we find that none of the results change by more than their 1-$\sigma$ uncertainties as long as the reduced X-ray data cover at least half of the original radial range.  We therefore assume that our results our reliable to a radius a factor of two beyond the outermost X-ray annulus.  In \Fref{tab:cM_constraints} we only quote contraints at a given overdensity radius $r_{\Delta \mbox{\tiny ref}}$ for those clusters whose X-ray data extends past $\frac{1}{2} r_{\Delta \mbox{\tiny ref}}$.  The same criteria is used to determine what clusters are included in the combined constraint on the nonthermal pressure fraction presented in \Fref{tab:nonthermal_pressure_fraction}.

In order to test the robustness of our result to the particular parameterization of the nonthermal pressure fraction profile, we have repeated the above analysis using a simple piecewise linear function
\[ \mathpzc{F}\subsc{outer}(r) = 
\begin{cases}
a + b \left(\frac{r}{r_{200m}}\right) &  r < r_{200m} \\
a + b & r \geq r_{200m}
\end{cases}
\]
with both the intercept $a$ and slope $b$ allowed to vary.  A uniform prior $U(0,1)$ is placed on both $a$ and $b$, resulting in a nonthermal pressure fraction that linearly increases with radius until $r_{200m}$ and is constant thereafter.  This model has one more parameter than the simulation-based model and allows for greater freedom in the shape of the profile.  We must, however, correct for the fact that the implicit prior on the nonthermal pressure fraction at a particular radius is nonuniform and radially dependent.  This is accomplished by dividing the measured posterior distribution for the nonthermal pressure fraction at the radius of interest by the analytical expression for the implicit prior, assuming our best-fit estimate of that radius and $r_{200m}$.  After making this correction we find nearly identical constraints as those obtained with the simulation-based model.

We have also derived frequentist confidence intervals on the normalization $C$ using the following method.  We step over a grid of $C$ values between $0$ and $1.825$.  At each point in the grid, we fix $C$ to the same value for all five clusters and use JACO to find the minimum $\chi^2$ allowing the other parameters of the model to float.  We then sum over the five clusters and examine $\left(\sum\chi^2\right)(C)$.  We find that the minimum value of $\sum\chi^2$ occurs at $C=0$.  We obtain a $95\%$ confidence interval by determining the value of $C$ where $\left(\sum\chi^2\right)(C) - \left(\sum\chi^2\right)(0) = 4.0$.  This results in $C = (0, \ 0.35)$, which is similar to the constraints obtained with the Bayesian approach.

\subsection{Abell 383}
\label{sec:abell_383}

We now discuss each cluster individually, starting with Abell 383.  This is the closest cluster in our sample at a redshift of $z=0.188$ and has relatively high-quality X-ray data.  The best fit to XR has a PTE of 0.00086.  This cluster has two independent measurements of the X-ray spectrum in each annular bin from the ACIS-I and ACIS-S imaging spectrometers.  The poor XR quality of fit is driven primarily by differences in these two measurements in three of the annular bins: the two innermost bins and the outermost bin.

As mentioned in the previous section, the mass inferred from GL is $1.6$--$2$ times larger than the mass inferred from XR or XR+SZ.  In addition, the X-ray and SZ data disagree with one another.  The SZ signal predicted from the XR-determined pressure is systematically lower than what is actually observed in the region between $200$ and $500 \ \mbox{kpc}$.  The X-ray data dominates the XR+SZ+GL fit, and hence underestimation of both the SZ and lensing signal by the best-fit model is apparent in \Fref{fig:radial_jaco_fit1}. 

These results further support the idea that Abell 383 is elongated along the line-of-sight direction \citep{newman:2011,morandi:2012}.  Such a geometry would naturally produce the discrepancies observed in our spherical fits to X-ray, SZ, and lensing data.  The equation of hydrostatic equilibrium implies that the ICM ``follows'' the gravitational potential.  More specifically, surfaces of constant gas density (and pressure) coincide with surfaces of constant gravitational potential.  A consequence of the Poisson equation is that the gravitational potential is more spherical than the density field that sources it.  Therefore, the gas density will in general be more spherical than the total density in dynamically relaxed galaxy clusters.  The X-ray and SZ observables are proportional to the gas density projected along the line of sight, whereas the lensing observable is proportional to the total density projected along the line of sight.  Elongation of the cluster along the line of sight will be more pronounced in the total density than the gas density, and will therefore result in a larger lensing signal than what is predicted based on either SZ or X-ray.  In addition, elongation will result in a larger SZ signal than what is predicted from the X-ray because the SZ observable scales as $\rho\subsc{gas}$ whereas the X-ray observable scales as $\rho\subsc{gas}^2$.

\citet{newman:2011} combined X-ray mass estimates with HST strong lensing data, Subaru weak lensing data, and measurements of the brightest cluster galaxy (BCG) stellar velocity dispersion profile to constrain a triaxial gNFW model for the dark matter halo assuming a major axis oriented along the line of sight.  The X-ray mass estimates were derived assuming spherical symmetry and hydrostatic equilibrium with a constant $10\%$ nonthermal pressure fraction, and were taken to represent the true, spherically averaged three-dimensional mass.  The projected mass profile measured by the lensing data was then used to constrain the line-of-sight extent of the dark matter halo $\eta\subsc{DM,a}^{-1} = 1.97^{+0.28}_{-0.16}$.  \citet{morandi:2012} performed a joint analysis of \emph{Chandra} X-ray and HST strong lensing data in which they fit a fully triaxial model for the dark matter and gas distribution.  They found that the data was well described by a triaxial dark matter halo with axis ratios $\eta\subsc{DM,a} = 0.55 \pm 0.06$ (minor/major) and $\eta\subsc{DM,b} = 0.71 \pm 0.10$ (intermediate/major) with the major axis of the dark matter halo inclined $21.1^{\circ} \pm 10.1^{\circ}$ from the line-of-sight direction.  They also included a constant nonthermal pressure fraction in their model and obtained the constraint $\mathpzc{F} = 0.11 \pm 0.05$.  Both of these works suggest that Abell 383 has a line of sight extent that is roughly a factor of 2 larger than its extent in the plane of sky.

In sum, both our analysis and previous works indicate that Abell 383 is poorly described by a spherical model. Our results for this cluster should therefore be considered with caution, and we forgo any detailed comparisons to other previous works based on a spherical analysis.

\subsection{Abell 611}
\label{sec:abell_611}

The primary peak of the convergence map is offset from that of the X-ray emission for Abell 611.  This results in a slightly lower concentration from the lensing-only fit than that found in \citet{merten:2015}.  The effect on the multiwavelength analysis is negligible.  The multiwavelength data is in good agreement with a spherical model with completely thermal pressure support.  This places a significant upper bound on the nonthermal pressure fraction.

Our finding that Abell 611 is approximately spherical is in good agreement with most previous results. For example, \citet{newman:2013} included it in their relaxed sample of seven clusters used to study cluster mass profiles and it would have been included in the relaxed sample defined by \citet{mantz:2015a} had it not failed their X-ray peakiness criteria. In addition, \citet{donnarumma:2011} derive consistent masses for Abell 611 using both X-ray and strong lensing data, further indicating that it is approximately spherical. However, we note that \citet{romero:2016} find, at a significance of $1.6\sigma$, evidence for an elongation along the line of site when comparing Bolocam and MUSTANG SZ data with \emph{Chandra} X-ray observations.

Abell 611 has been the focus of a wide range of lensing analyses, with \citet{applegate:2016} and \citet{hoekstra:2015} finding values of $M_{\mbox{\footnotesize tot,} 2500c}$ that are consistent with our results at the $\simeq 1 \sigma$ level. However, both \citet{okabe:2016} (at $M_{\mbox{\footnotesize tot,} 500c}$) and \citet{newman:2013} (at $M_{\mbox{\footnotesize tot,} 200c}$) obtain lensing-derived masses approximately $2 \sigma$ lower than our results, and the value of $M_{\mbox{\footnotesize tot,} 200c}$ obtained by \citet{romano:2010} is less than half of our value. However, the X-ray hydrostatic value of $M_{\mbox{\footnotesize tot,} 2500c}$ derived by \citet{applegate:2016} is consistent with our results. While that lack of overall agreement in these results is somewhat concerning, we emphasize our consistency with \citet{applegate:2016} and \citet{hoekstra:2015}, both of which were focused on accurate mass calibration of large cluster samples to enable precise cosmological studies.

\subsection{MACS J0429.6-0253} 

MACS J0429.6-0253 also has an offset between the X-ray- and lensing-determined centers.  The net result is the same as in Abell 611.  There is slight tension between the X-ray and SZ data.  This manifests as an excess in the measured SZ signal over what is expected based on the XR determined pressure in the region between $500$ and $900 \ \mbox{kpc}$.  This difference is not statistically significant, however, and our model is able to provide a good quality of fit.  

Several previous studies have found this cluster to be among the most relaxed objects in their samples \citep{maughan:2008, mann:2012, mantz:2015a}, although \citet{romero:2016} find that the overall normalizations of the X-ray and SZ signals differ at a significance of $1.5\sigma$. In addition, the total mass we obtain for MACS J0429.6-0253 is consistent with the lensing results of \citet{merten:2015} and \citet{applegate:2016}, along with the X-ray hydrostatic analysis of \citet{applegate:2016}, further demonstrating the relatively relaxed dynamical state of this cluster.

\subsection{MACS J1311.0-0310 and MACS J1423.8+2404}

The multiwavelength data for these two clusters is well described by a spherical model with completely thermal pressure support.  However, the data are also consistent with a wide range of nonthermal pressure normalizations.  The lack of constraining power on the nonthermal pressure fraction is likely due to the relatively high redshifts of these clusters, which results in X-ray data that only extends out to $0.7 \times r_{500c}$ in the case of MACS J1311.0-0310 and $0.3 \times r_{500c}$ in the case of MACS J1423.8+2404.  The X-ray data is necessary to constrain the gas density, which can be degenerate with the nonthermal pressure fraction.

Most previous studies have similarly found these clusters to be highly relaxed and approximately spherical \citep{maughan:2008, mann:2012, mantz:2015a}, with \citet{adam:2016} also finding good agreement between X-ray data from XMM and SZ data from NIKA for MACS J1423.8+2404. However, as with the previous two clusters, \citet{romero:2016} find that the X-ray and SZ signals differ at modest significance. In addition, we note that MACS J1311.0-0310 was given a slightly elevated morphological classification by \cite{mann:2012} compared to the most relaxed objects, and \citet{limousin:2010} found slight tension between X-ray and lensing data for MACS J1423.8+2404, suggestive of a line of sight elongation.  Our derived masses for these clusters are in good agreement with previous lensing \citep{merten:2015, applegate:2016} and X-ray hydrostatic measurements \citep{bonamente:2008, applegate:2016, adam:2016}, although we do note that our mass for MACS J1423.8+2404 is a little more than $1\sigma$ higher than the X-ray hydrostatic mass found by \citet{bonamente:2008}.

\subsection{MACS J1532.8+3021}
\label{sec:macs_j1532}

No strong lensing features were found in the HST data for MACSJ1532.8+3021, making it the only cluster in our sample without gravitational strong lensing constraints.  It does have high-quality X-ray data that extends out to $r_{500c}$.  The X-ray data dominates the multiwavelength fits.  The X-ray and SZ data agree remarkably well outside of $\sim 400 \ \mbox{kpc}$.  Within this radius, however, there is a significant discrepancy between the X-ray and SZ data.  This cluster contains a powerful AGN that is almost certainly responsible for the disagreement in the cluster core \citep{hlavacek-larrondo:2013}.  In our multiwavelength analysis, this results in a significant inner nonthermal pressure component $\mathpzc{F}\subsc{inner}$ that approaches $\simeq 28\%$ as $r \rightarrow 0$.  This is the only cluster in our sample where the $F$-test prefers an inner nonthermal pressure component.

Based on other published results, all of which indicate this cluster is highly relaxed, the significant amount of nonthermal pressure is somewhat surprising \citep{maughan:2008, mann:2012, mantz:2015a}. However, these previous studies were all based on X-ray morphology, and were therefore insensitive to the effects of nonthermal pressure. In addition, the mass we obtain for MACS J1532.8+3021 is approximately 25\% lower compared to both the lensing and X-ray hydrostatic measurements of \citet{applegate:2016}, although, in the case of the lensing mass, the statistical significance of the disagreement is modest ($\simeq 1\sigma$). Further, we note that our X-ray-only hydrostatic mass is in good agreement with the measurements of \citet{applegate:2016}, indicating that the expanded dataset and/or model parameters used in our analysis are the likely cause of the difference.

\section{Discussion}
\label{sec:jaco_discussion}

The multiwavelength analysis results in significant improvement in the constraints on both the concentration and mass of the five galaxy clusters examined.  First, comparing the XR analysis to the XR+SZ analysis, there is a median reduction of $8\%$ in the uncertainty on the concentration and $35$--$40\%$ in the uncertainty on the total mass over the radial range $r_{2500c}$--$r_{200c}$.  This type of joint X-ray and SZ analysis is well-suited for obtaining mass estimates for high-$z$ clusters, where deep X-ray observations are expensive due to cosmological dimming.  Next, comparing the XR+SZ analysis to the XR+SZ+GL analysis, we find that the median reduction in the uncertainty on the mass and concentration is minimal, at the $0$--$10\%$ level.  However, the addition of lensing data allows us to examine whether nonthermal pressure support is necessary to describe the cluster, and if so, include it in our model, thereby mitigating this known systematic bias in the resulting mass estimate.  Finally, comparing the GL analysis to the XR+SZ+GL analysis, we see a dramatic improvement in the constraints on both the concentration and mass.  There is a median reduction of $50$--$55\%$ in the uncertainty on the concentration and $50$--$70\%$ in the uncertainty on the total mass over the radial range $r_{2500c}$--$r_{200c}$.  This results in a $80$--$85\%$ reduction in the area of the $68\%$ and $95\%$ concentration-mass credible regions.  

Compared to hydrodynamical simulations, we observe a distinct lack of nonthermal pressure support in the subset of five galaxy clusters.  We now discuss assumptions implicit to our analysis that may effect these results.  
\begin{itemize}
\item We do not include systematic errors for the possible miscalibration of X-ray temperatures.  \citet{donahue:2014} performed a comparison of the density and temperature profiles derived from \emph{Chandra} and XMM data for the X-ray-selected sample of CLASH clusters.  They found that the gas density profiles measured by the two instruments were in excellent agreement.  The temperature profiles were also in good agreement in the cluster core.  However, the XMM temperatures systematically declined relative to the \emph{Chandra} temperatures with increasing radius.  If \emph{Chandra} overestimates the gas temperature, then that would result in a underestimation of the level of nonthermal pressure support.  However, in our analysis the gas temperature constraints at large radius are driven primarily by the SZ data, thereby mitigating any potential errors related to X-ray calibration in the regions where \emph{Chandra} and XMM are most discrepant.
\item Our model assumes spherically symmetry; however, there is significant evidence from both observation \citep{carter:1980,binggeli:1982,fabricant:1984,buote:1992,evans:2009,kawahara:2010,oguri:2010,sayers:2011,oguri:2012} and numerical simulation \citep{frenk:1988,dubinski:1991,warren:1991,jing:2002,hopkins:2005,meneghetti:2010,munoz-cuartas:2011,lemze:2012,limousin:2013,despali:2013} that galaxy clusters are better approximated as triaxial ellipsoids.  Departures from sphericity could potentially bias our estimate of the nonthermal pressure fraction due to the degenerate manner in which it affects the multiwavelength observables.  Specifically, both nonthermal pressure and line-of-sight elongation (or compression) result in differences between the lensing- and X-ray/SZ-inferred mass.  We have selected clusters that are circular in the plane of the sky as evidenced by both X-ray and SZ data.  Therefore, any departure from sphericity would have to occur primarily along the line-of-sight direction.  If the clusters were elongated along the line-of-sight, our analysis would overestimate the level of nonthermal pressure support, similar to what is seen in Abell 383.  Only a compression of the galaxy cluster along the line-of-sight would result in the underestimation of the nonthermal pressure support necessary to explain the discrepancy between our results and hydrodynamical simulations:  the galaxy clusters would have to be oblate ellipsoids with minor axis oriented along the line-of-sight direction.  However, this geometry would also result in discrepancies between the X-ray and SZ data under the spherical model, with the pressure inferred from the X-ray data predicting more SZ signal than what is actually observed.  We do not see this type of behavior in any of the clusters.  In general, other than the modest tension observed in MACS J0429.6-0253 and MACS J1532.8+3021, there is good agreement between the Bolocam SZ and \emph{Chandra} X-ray data in the regions of radial overlap for these five clusters.  We are unable at this time to make this constraint on oblateness quantitative, though we will in future, fully triaxial analyses.
\item Selection effects may also be responsible for the lack of nonthermal pressure support.  The X-ray selected sample of 20 CLASH clusters were chosen to exhibit a high degree of regularity based on \emph{Chandra} images of the X-ray surface brightness \citep{postman:2012,meneghetti:2014}.  In this work, we further restricted our attention to the 6 clusters in the sample with the most regular morphology by placing cuts on the X-ray centroid shift parameter $w$ and the ellipticity of the SZ image.  Therefore, our analysis is focused on a very distinct type of cluster, namely ones with gas density and pressure distributions that are azimuthally symmetric when projected onto the plane of the sky.  Finally, by discarding Abell 383 from the nonthermal pressure analysis, we further restricted our attention to clusters whose X-ray and SZ data do not show show evidence of line-of-sight elongation.
\end{itemize}

The galaxy cluster Abell 383 illustrates the fact that even the most regular clusters can have significant elongation along the line-of-sight.  Previous multiwavelength analyses of Abell 383 employed X-ray and lensing data \citep{newman:2011,morandi:2012}.  The new insight gained from our analysis is that the SZ signal scales in the expected way for line-of-sight elongation.  The tension between the X-ray and SZ can therefore be used to constrain the line-of-sight extent, breaking the degeneracy that exists between line-of-sight extent and nonthermal pressure support.  Triaxial modeling of the dark matter and ICM mass distributions is necessary to do this properly.  Future work will be directed towards generalizing the JACO software to fit triaxial models and performing a triaxial analysis of the entire sample of 25 CLASH clusters.

\section{Summary}
\label{sec:jaco_summary}

We have performed a multiwavelength analysis of a subset of 6 of the 20 X-ray selected CLASH clusters.  The subsample was selected by placing a stringent cut on the X-ray centroid shift parameter $w$ derived from \emph{Chandra} measurements of the X-ray surface brightness and also placing a cut on the ellipticity of the SZ image measured by Bolocam.  These criteria select clusters with gas density and pressure distributions that are azimuthally symmetric when projected onto the plane of the sky.  For each cluster, the JACO software was used to fit a parametric model to a set of radially binned X-ray spectra measured by \emph{Chandra}, a radially binned convergence profile derived from HST/Subaru strong and weak lensing data, and a two-dimensional SZ image measured by Bolocam.  A statistical $F$-test was employed to determine the maximally restricted model necessary to describe the data.  Various subsets of the multiwavelength data were fit to understand the relative contribution to the resulting constraints on the concentration and total mass of the galaxy cluster.

We find that, for 5 of the 6 clusters, a relatively simple model that assumes spherical symmetry, hydrostatic equilibrium, and entirely thermal pressure support provides a good fit to the multiwavelength dataset.  There are significant improvements $(35$--$40\%)$ in the constraints on the total mass obtained from the joint fits to the X-ray and SZ data relative to those obtained from fits to the X-ray data only.  There are also significant improvements in the constraints on both the concentration $(50$--$55\%)$ and total mass $(50$--$70\%)$ obtained from the joint fits to the X-ray, SZ, and lensing data relative to those obtained from fits to the lensing data only.  

The five clusters that are well described by the model are used to place an upper bound on the level of nonthermal pressure support present in the ICM.  We find that the nonthermal pressure at $r_{500c}$ is less than $11\%$ of the total pressure at $95\%$ confidence.  This is in tension with state-of-the-art hydrodynamical simulations, which suggest nonthermal pressure fractions of $26\%$ at $r_{500c}$ for clusters of this mass and redshift.  Possible causes for this discrepancy include selection effects, X-ray temperature miscalibration, and compression of the cluster along the line-of-sight direction.

We find that, for one of the clusters, Abell 383, the multiwavelength data disagrees in a way that suggests the cluster is elongated along the line-of-sight direction.  Future work will generalize the model to allow the cluster to have a triaxial shape and arbitrary orientation.  In addition, the software will be upgraded to fit X-ray surface brightness images in addition to radially binned spectra.   Currently the outer radius that we can reliably constrain the model is limited by the extent of the X-ray spectra, and including the surface brightness in the fit will extend results to larger radii.  After these modifications, the analysis will be expanded to the entire sample of 25 CLASH clusters.

\section{Acknowledgments}
\label{sec:acknowledgments}

We acknowledge the assistance of: the day crew and Hilo staff of the Caltech Submillimeter Observatory, who provided invaluable assistance during commissioning and data-taking for the Bolocam dataset; Kathy Deniston, Barbara Wertz, and Diana Bisel, who provided effective administrative support at Caltech and in Hilo.  This research made extensive use of high performance computing resources at the Caltech Center for Advanced Computing Research and the SFSU Facility for Space and Terrestrial Advanced Research.  SRS was supported by NASA Earth and Space Science Fellowship NASA/NNX12AL62H and a generous donation from the Gordon and Betty Moore Foundation.  JS was supported by NSF/AST1313447, NASA/ NNX11AB07G, and the Norris Foundation CCAT Postdoctoral Fellowship.  AZ is supported by NASA through Hubble Fellowship grant \#HST-HF2-51334.001-A awarded by STScI, which is operated by the Association of Universities for Research in Astronomy, Inc. under NASA contract NAS~5-26555.  JM has received funding from the
People Programme (Marie Curie Actions) of the European Unions Seventh Framework Programme (FP7/2007-2013) under REA grant agreement number 627288.

\emph{Facilities:}  Caltech Submillimeter Observatory, \emph{Chandra}, Hubble Space Telescope, Subaru

\bibliographystyle{apj}
\bibliography{library}


\begin{appendix}
\section{Comparison of Data and Best-fit Maximally Restricted Model}
\label{sec:best_fit_model_figs}
\begin{figure*}
    \centering
	\includegraphics[width=\linewidth,keepaspectratio]{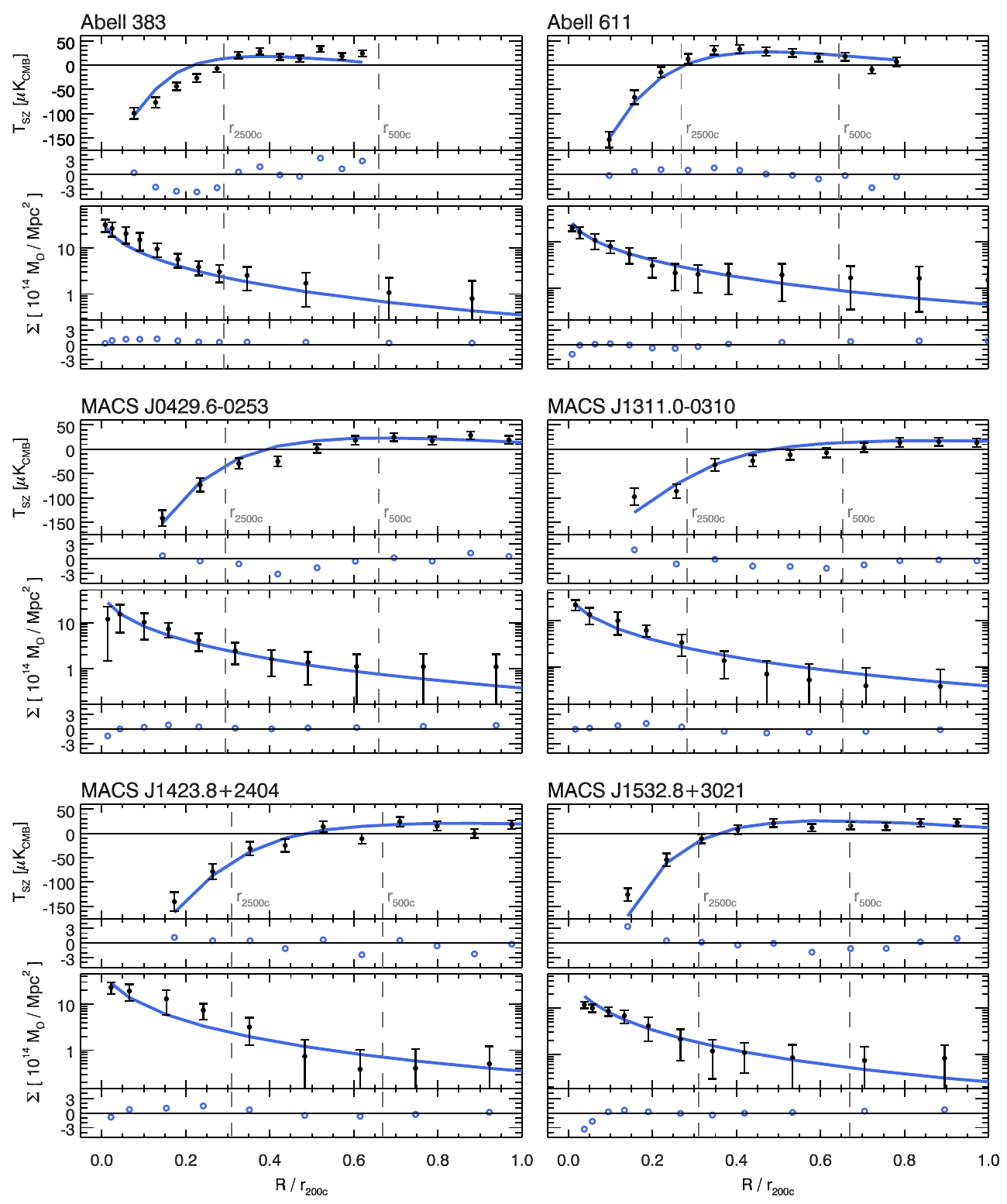}
	\caption{The SZ and lensing data compared to the best-fit, maximally restricted model as a function of projected radius relative to $r_{200c}$.  Note that the model is determined from a fit to the full multiwavelength dataset.  The top portion of each panel displays the data and best-fit model.  The bottom portion of each panel displays the normalized residuals (i.e., $[\mbox{data} - \mbox{model}] / \mbox{uncertainty}$).  The upper panel is the azimuthally averaged SZ image.  Note that we fit the two-dimensional SZ image, but show the radial profile here for visualization purposes.  Recall that the SZ effect results in a temperature decrement, and the positive excursion at intermediate radii is due to the filtering applied during data processing.  The bottom panel is the convergence profile reconstructed by the \texttt{SaWLens} algorithm from both strong and weak lensing constraints, converted to a surface mass density profile $\Sigma = \kappa \Sigma\subsc{crit}$.  The X-ray data are presented on the following pages.} 
	\label{fig:radial_jaco_fit1}
\end{figure*}

\begin{figure*} 
    \centering
	\includegraphics[width=\linewidth,keepaspectratio]{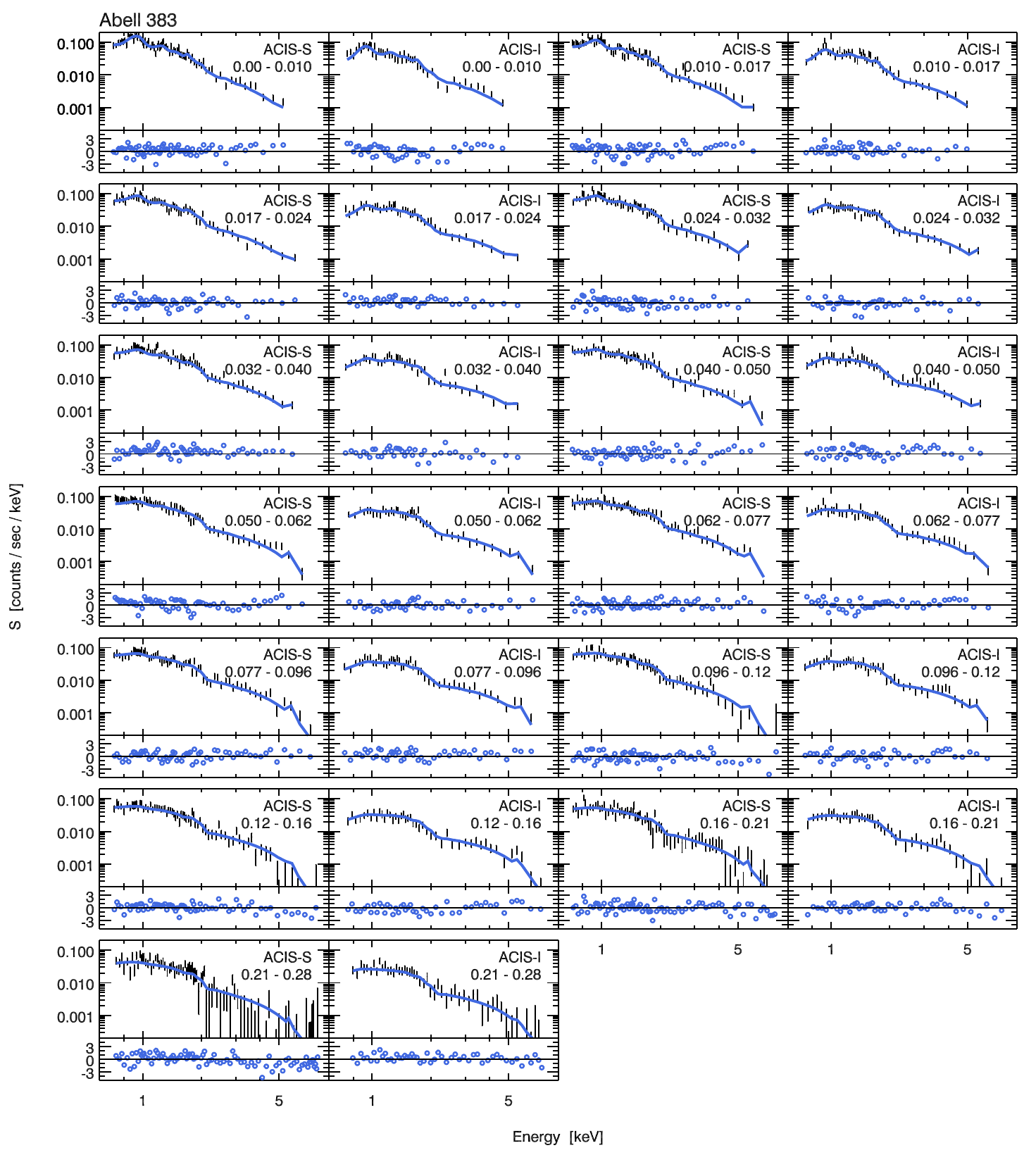}
	\caption{The measured cluster X-ray spectra compared to the best-fit, maximally restricted model for Abell~383.  Note that the model is determined from a fit to the full multiwavelength dataset.  Each panel is a different detector/annular bin.  The detector name and the inner and outer radii relative to $r_{200c}$ are annotated in the upper right corner.  The top portion of each panel displays the data and best fit model.  The bottom portion of each panel displays the normalized residuals (i.e., $[\mbox{data} - \mbox{model}] / \mbox{uncertainty}$).} 
	\label{fig:radial_jaco_fit2}
\end{figure*}

\begin{figure*}
    \centering
	\includegraphics[width=\linewidth,keepaspectratio]{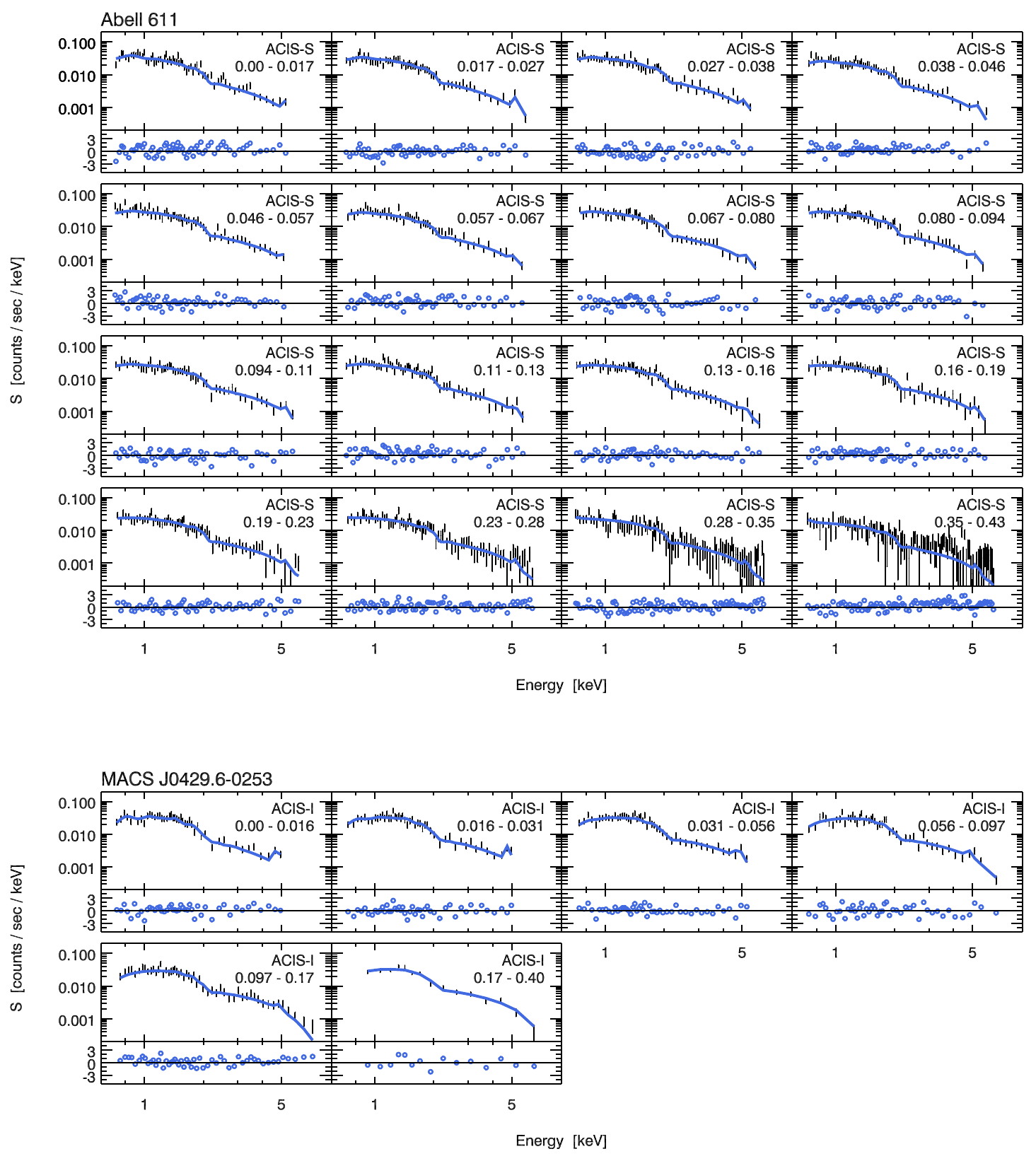}
	\caption{The measured cluster X-ray spectra compared to the best-fit, maximally restricted model for Abell~611 and MACS~J0429.6-0253.  Note that the model is determined from a fit to the full multiwavelength dataset.  Each panel is a different detector/annular bin.  The detector name and the inner and outer radii relative to $r_{200c}$ are annotated in the upper right corner.  The top portion of each panel displays the data and best fit model.  The bottom portion of each panel displays the normalized residuals (i.e., $[\mbox{data} - \mbox{model}] / \mbox{uncertainty}$).} 
	\label{fig:radial_jaco_fit3}
\end{figure*}

\begin{figure*}
    \centering
	\includegraphics[width=\linewidth,keepaspectratio]{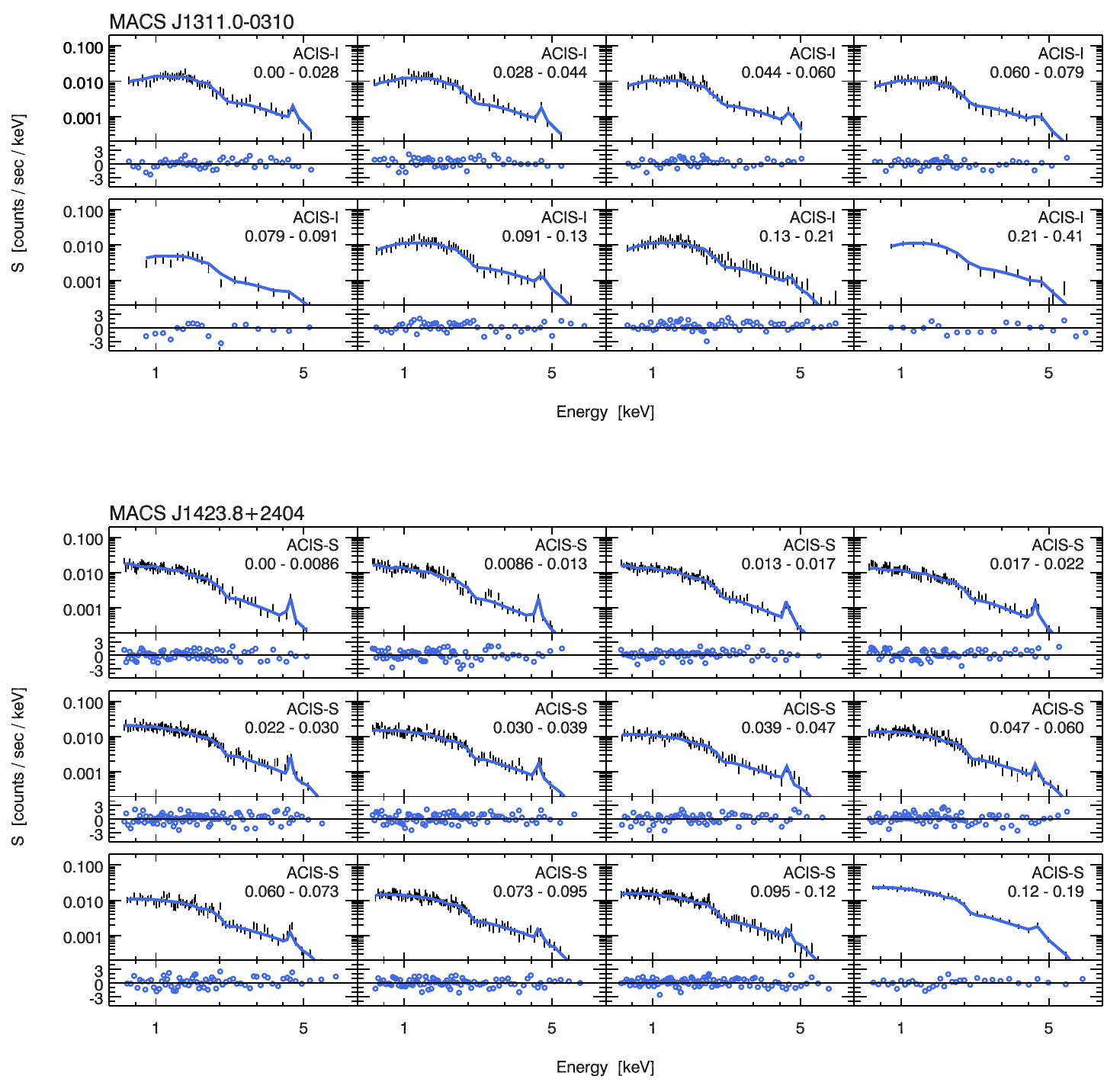}
	\caption{The measured cluster X-ray spectra compared to the best-fit, maximally restricted model for MACS~J1311.0-0310 and MACS~J1423.8+2404.  Note that the model is determined from a fit to the full multiwavelength dataset.  Each panel is a different detector/annular bin.  The detector name and the inner and outer radii relative to $r_{200c}$ are annotated in the upper right corner.  The top portion of each panel displays the data and best fit model.  The bottom portion of each panel displays the normalized residuals (i.e., $[\mbox{data} - \mbox{model}] / \mbox{uncertainty}$).} 
	\label{fig:radial_jaco_fit4}
\end{figure*}

\begin{figure*}
    \centering
	\includegraphics[width=\linewidth,keepaspectratio]{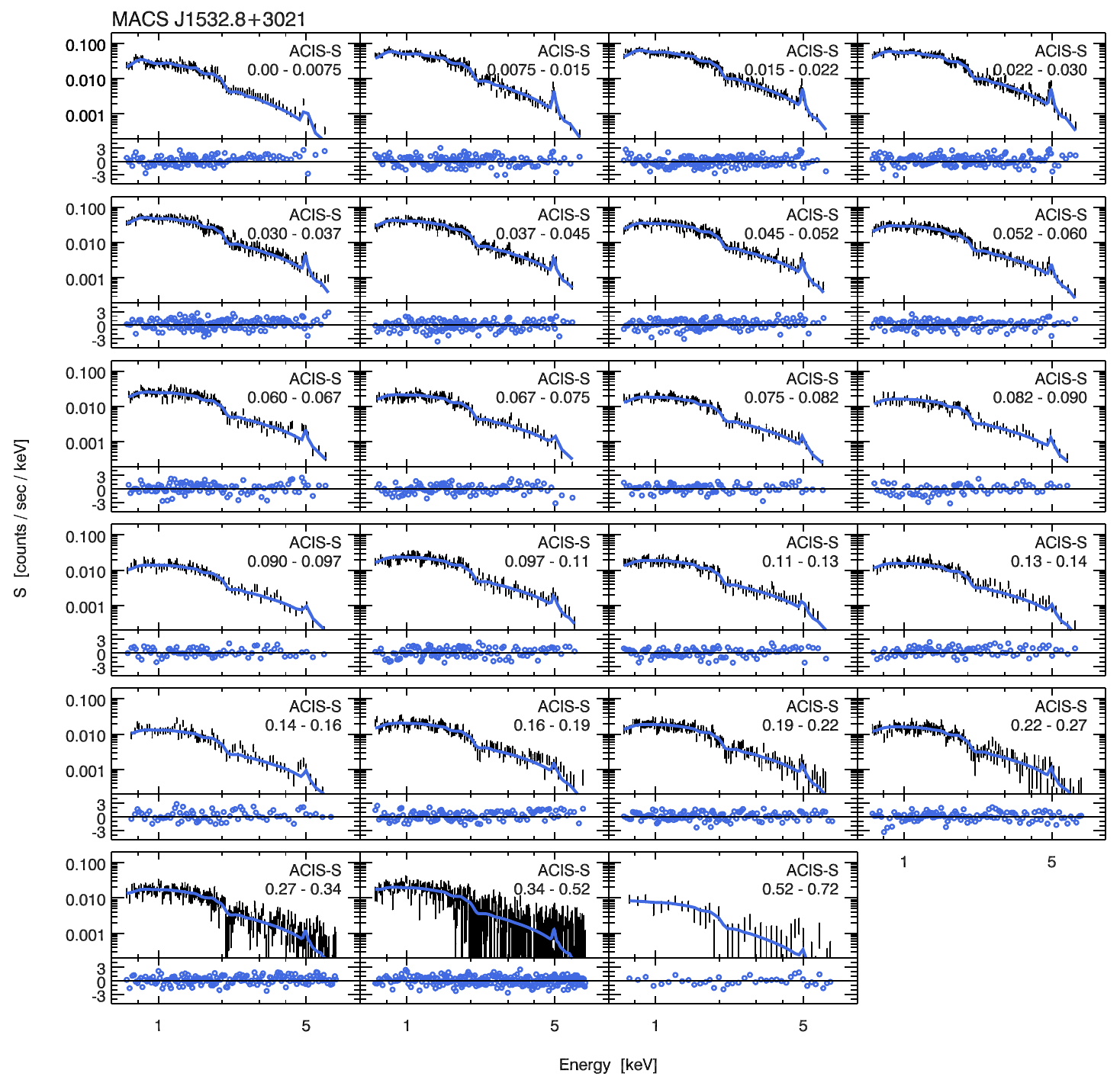}
	\caption{The measured cluster X-ray spectra compared to the best-fit, maximally restricted model for MACS~J1532.8+3021.  Note that the model is determined from a fit to the full multiwavelength dataset.  Each panel is a different detector/annular bin.  The detector name and the inner and outer radii relative to $r_{200c}$ are annotated in the upper right corner.  The top portion of each panel displays the data and best fit model.  The bottom portion of each panel displays the normalized residuals (i.e., $[\mbox{data} - \mbox{model}] / \mbox{uncertainty}$).} 
	\label{fig:radial_jaco_fit5}
\end{figure*}

\end{appendix}
	
\end{document}